\documentclass[twocolumn]{autart}

\usepackage[gen]{eurosym}
\usepackage{dsfont}

\usepackage{multicol}
\usepackage{enumerate}
\usepackage{balance}

\usepackage{mathrsfs}
\usepackage[bb=boondox]{mathalfa}
\usepackage{subfigure}
\usepackage{graphicx,color}
\usepackage{latexsym}
\usepackage{amsmath} 
\usepackage{amssymb}  
\usepackage{amsmath,amssymb,amsfonts,theorem}
\usepackage{pgf}
\usepackage{tikz}
\usepackage{circuitikz}
\usetikzlibrary{arrows,shadows,calc,decorations.shapes,decorations, automata} 
\usepackage[underline=true,rounded corners=false]{pgf-umlsd}
\usepackage{cuted}
\usepackage{dsfont}

\usepackage{bbold}
\newtheorem{remark}{Remark}
\newtheorem{definition}{Definition}

\newtheorem{theorem}{Theorem}
\newtheorem{lemma}{Lemma}

\newtheorem{assumption}{Assumption}
\newtheorem{corollary}{Corollary}
\newtheorem{problem}{Problem}
\newtheorem{objective}{Objective}
\newtheorem{constraint}{Constraint}
\def\be{\begin{equation}}
\def\ee{\end{equation}}
\def\ba{\begin{array}}
\def\ea{\end{array}}

\def\stopmodif{\color{black}}

\def\qedp{\hspace*{\fill}~{\tiny $\blacksquare$}}

\newcommand{\R}{\mathds{R}}

\def\s1{\color{purple}}
\def\e1{\color{black}}

\usepackage{natbib}
\begin{document}
\begin{frontmatter}

\title{ Optimal regulation of  flow networks with transient constraints$^{\star}$}

\thanks[footnoteinfo]{This work is partially supported by the Danish Council for Strategic Research (contract no. 11-116843) within the `Programme Sustainable Energy and Environment', under the `EDGE' (Efficient Distribution of Green Energy) research project and by the research grant `Flexiheat' (Ministerie van Economische Zaken, Landbouw en Innovatie). Preliminary results have appeared in \cite{trip_2017_ifac}.\\
$^\diamond$ Both authors contributed equally.}

\author[First]{Sebastian Trip$^\diamond$}
\author[First,Second]{$\quad$Tjardo Scholten$^\diamond$}
\author[First]{$\quad$Claudio De Persis}

\address[First]{ENTEG, Faculty of Science and Engineering, University of Groningen, Nijenborgh 4, 9747 AG Groningen, the Netherlands. (e-mail: s.trip@rug.nl;  t.w.scholten@rug.nl; c.de.persis@rug.nl). }
\address[Second]{Institute of Engineering, Hanze University of Applied Sciences,  Zernikeplein 11,  9747 AS Groningen, the Netherlands.}

\begin{abstract}   This paper investigates the control of flow networks, where the control objective is to regulate the measured output (e.g storage levels) towards a desired value.  We present a distributed controller that dynamically adjusts the inputs and flows, to achieve output regulation in the presence of unknown disturbances, while satisfying given input and flow constraints. Optimal coordination among the inputs, minimizing a suitable cost function, is achieved by exchanging information over a communication network.  Exploiting an incremental passivity property, the desired steady state is proven to be globally asymptotically attractive under the closed loop dynamics. Two case studies (a district heating system and a multi-terminal HVDC network) show the effectiveness of the proposed solution.
\end{abstract}

\begin{keyword}
Control of networks, Optimization, Passivity, Distributed control.
\end{keyword}

\end{frontmatter}

\section{Introduction}
Flow networks (also known as distribution or transportation networks) consist of edges that are used to model the
exchange of material (flow) between the nodes.   The design and regulation of these networks received significant attention due to its many applications, including supply chains (\cite{alessandri_2011_tcst}), heating, ventilation and air conditioning (HVAC) systems (\cite{gupta2015distributed}), data networks (\cite{moss_1982_tac}), traffic networks (\cite{iftar_1999_automatica}, \cite{coogan_2015_tac}) and compartmental systems (\cite{blanchini_2016_automatica}, \cite{como_2017_arc}).
 If the considered objective is static, the study of flow networks has a long history within the field of network optimization (\cite{bertsekas1998network}, \cite{rockafellar1984network}). Many practical networks must on the other hand
 react dynamically on changes in the external conditions such as a change in the demand. In these cases continuous feedback controllers are required, that dynamically adjust inputs at the nodes and the flows along the edges, and the design of such controllers is the subject of this work.


Since flow networks are ubiquitous in engineering systems, many solutions have been proposed to coordinate them, exploiting methodologies from {\it e.g.} 
passivity (\cite{arcak2007}) and model predictive control (\cite{koeln_2017_automatica}).
We focus on flow networks where the nodes can store the considered material (\cite{kotnyek2003annotated}).
A common objective in such networks is that the stored material needs to be regulated towards desired setpoints, despite the presence of an unknown demand. This is commonly achieved by actively controlling the flows on the edges (\cite{wei2013load}, \cite{burger_2015_automatica}, \cite{xiang_2017_automatica}) using \emph{dynamic} flow controllers.
These controllers on the edges generally provide a form of integral action, that shows some benefits over networks lacking these dynamics. For example, the presence of an integral action permits the achievement of output regulation, in contrast to approximate regulation (\cite{giordano2016structural}). This inability to achieve output regulation in the presence of unknown disturbances can  be observed in systems where the flow on an edge proportionally depends on the potential difference of its adjacent nodes. This is found in {\it e.g.} compartmental systems (\cite{blanchini_2016_automatica} \cite{como_2017_arc}). Furthermore, in most cases, the capacity of the edges is constrained, requiring careful design of the flow controllers. Naturally, the control of flows only permits to distribute the material within the network. In case there is no possibility to adjust the input to the network, a necessary requirement for stability is that all uncontrollable inflows and outflows sum to zero (\cite{wei2016consensus}). Since this is generally not the case, additional controllable inputs are required that might have their own capacity constraints.

\subsection{Main contributions}
In this work we focus on flow networks, where at various nodes, an \emph{unknown} amount of material (disturbance) is supplied to, or extracted from, the network. Despite these disturbances, we require the various storage levels at the nodes (or an `output function' thereof) to be regulated towards desired values. We aim at achieving this so-called \emph{output regulation}, by \emph{optimally} allocating the required inputs among the nodes that possess a controllable external input.  Here, only a subset of the nodes is assumed to have a controllable input, where a cost function relates the provided input to associated costs. We particularly propose a \emph{distributed} control solution to enhance robustness to failures and to improve the scalability. Furthermore, the proposed solution respects capacity constraints that the inputs and flows might have.

Although various of these aspects  have been addressed before, the way how we incorporate  them within a coherent approach is new. Furthermore, the proposed controllers are shown to achieve the overall objective outlined above \emph{globally}, {\it i.e.} independent of the initialization of the system. We elaborate on some specific contributions below.

(i) In  flow networks it is desirable to meet certain optimality criteria, prescribing {\it e.g.} the optimal flows within the network and the optimal inputs to the network. Examples of the former include a `maximum flow', `quickest flow' or `minimum cost flow', and achieving them received a considerable amount of attention in the past  (see \cite{kotnyek2003annotated}, \cite{skutella2009introduction} and references therein). On the other hand, when optimal inputs are considered, costs are often associated to the amount of generated input (materials), and optimization thereof has been studied thoroughly within the setting of smart (electricity) grids (\cite{trip_2016_automatica}, \cite{dorfler_2016_tcns}). In this paper we apply this idea to general flow networks (\cite{scholten2016optimal}), where only a subset of the nodes can generate an input. A communication network then connects the various nodes, where relevant information on the costs is exchanged.

(ii) The distributed controllers are designed to enjoy certain passivity properties. That passivity plays an outstanding role in the coordination of systems is well recognized (\cite{arcak2007}). Particularly, incremental passivity (\cite{Pavlov2008}) has been exploited to analyze the stability of flow networks (see. {\it e.g.} \cite{burger_2015_tcns} and \cite{burger_2015_automatica}),  but also of virtual networks in the setting of distributed optimization (\cite{tang_2016_icca}) and game theory (\cite{gadjov_2017_arxiv}).
To prove asymptotic convergence to the desired state, generally, some form of strict output passivity ({\it e.g.} as a result of damping) is required. The considered flow networks in this work do not enjoy  this property, due to the preservation of the material, making the controller design more challenging. We propose a `dynamic extension' of previously considered integral-type controllers, to ensure convergence to a point, preventing the network to converge  to a limit cycle, exhibiting oscillations. Although the approach is tailored to the system at hand, the design offers new perspectives on similar systems lacking dissipation. In case physical considerations forbid this dynamic extension, global convergence to the desired output can be achieved by carefully selecting nodes that have a controllable input. This selection is related to the  zero forcing set of the underlying graph of the network (\cite{monshizadeh2014zero, trefois_2015_laa}), and this work provides an interesting link between zero forcing sets and the application of an invariance principle for dynamical systems.

  (iii) The proposed distributed controllers are applied, besides flow networks, to compartmental systems, studied  in {\it e.g.} \cite{blanchini_2016_automatica} and \cite{como_2017_arc}, and we show that additional control on some inputs and flows is sufficient to achieve regulation. Although setpoint regulation for (linear) compartmental systems has been studied before in \cite{lee_2015_tac} and \cite{ahn_2017_tc}, our approach is different. In the aforementioned works, the flows are adjusted by properly altering the system parameters of the network, whereas we consider here the parameters constant and dynamically adjust the flows on some edges that are independent of the state of the network.
    
   (iv) We provide two case studies that exemplify the use of flow networks to describe interconnected physical systems. In the first case study, we consider district heating systems (\cite{Scholten2015}) and improve upon existing results by guaranteeing asymptotic convergence to a desired setpoint, where only a subset of the nodes are required to have a controllable input. In the second case study, we consider voltage regulation and current sharing in multi-terminal high voltage direct current (HVDC) networks (\cite{zonetti_2015_cep}, \cite{andreasson_2016_tns}). Despite the fact that these networks have already been studied extensively, the proposed control solution is noteworthy in that it provides means to limit current injections during transients and does not require all terminals to be controlled.

\subsection{Outline}
The paper is structured as follows. In Section \ref{flow_networks} we introduce the considered flow network model. Next, in Section \ref{optimal_regulation}, we state our control objective of \emph{optimal output regulation} and discuss various constraints under which the control objective should be achieved. In Section \ref{controller_design} we propose a distributed controller and study the feasibility of the control problem in more detail. Exploiting incremental passivity properties of the network and the controllers, the stability analysis of the closed loop system is carried out in Section \ref{stability_analysis}. In Section \ref{section:physicalflows}, we study two modifications to the controlled flow network, widening the scope of this work. Two case studies are presented in Section \ref{case_study}. Finally, the conclusions and future directions are given in Section \ref{conclusions}.

\subsection{Notation}

Let $\mathbf{0}$ be the vector of all zeros of suitable dimension and let $\mathds{1}_n$ be the vector containing all ones of length $n$. The $i$-th element of vector $x$ is denoted by $x_i$ or, if it enhances the readability, by $[x]_i$. We define $\mathcal{R}(f)$ to be the range of function $f(x)$. A steady state solution to system $\dot x = f(x)$, is denoted by $\overline x$, {\it i.e.} $\boldsymbol{0} = f(\overline x)$. In case the argument of a function is clear from the context, we occasionally write $f(x)$ as $f(\cdot)$. Let $A \in \mathds{R}^{n \times m}$ be a matrix, then $\text{im}(A)$ is the image of $A$ and $\text{ker}(A)$ is the kernel of $A$. In case $A$ is a positive definite (positive semi-definite) matrix, we write $A \in \mathds{R}_{>0}^{n \times n}$  ($A \in \mathds{R}_{\geq0}^{n \times n}$). Lastly, we denote the cardinality of a set $\mathcal{V}$ as $|\mathcal{V}|$.

For convenience we provide, in Table \ref{tab:symbols}, an overview of some important symbols appearing in this work.
\begin{table}
\centering
{\begin{tabular}{ll}
	\bf{Symbol}										&\bf{Description}\\		
\hline			\\
$\mathcal{G}$ & Graph of the network \\
$\mathcal{V}$ & Set of nodes \\
$\mathcal{V}_e$ & Set of nodes with controllable external input\\
$\mathcal{E}$ & Set of edges \\
$B$ & Incidence matrix of the network \\
$E$ & Indicator matrix of controllable external inputs \\
$T_{\star}$ & Constant (gain) matrix \\
$L^{com}$ & Laplacian matrix of the communication graph \\
$Q$ & Quadratic cost matrix \\
$r$ & Linear cost vector \\
$x$ & Storage / inventory level \\
$y$ & Output ($y = h(x)$)\\
$\overline y$ & Desired output \\
$d$ & Disturbance / demand \\
$u$ & Controllable external input ($u = g(\theta)$) \\
$\overline u$ & Optimal input \\
$\lambda$ & Flows on the edges ($\lambda = f(\mu)$) \\
%
$\xi$ & Auxiliary flow controller state \\
$\phi$ & Auxiliary input controller state \\
\\
\end{tabular}}
\caption{Description of various symbols. }\label{tab:symbols}
\end{table}
\stopmodif
\section{Flow networks} \label{flow_networks}
In this paper we consider a network of physically interconnected \emph{undamped} dynamical systems. The topology of the system is described by an undirected graph $\mathcal{G}=(\mathcal{V},\mathcal{E})$, where $\mathcal{V} = \{1,...,n\}$ is the set of nodes and $\mathcal{E} 
= \{1,...,m\}$ is the set of edges connecting the nodes.
We represent the topology by its corresponding incidence matrix $B \in \R^{n \times m}$, where the entries of $B$ are defined by arbitrarily labelling the ends of the edges in $\mathcal{E}$ with a \lq +' and a \lq --', and letting
\begin{equation*}
\label{eq:incidence}
b_{ik}=
\begin{cases}
+1 \quad &\text{if node $i$ is the positive end of edge $k$}\\
-1 \quad &\text{if node $i$ is the negative end of edge $k$}\\
0 \quad &\text{otherwise}.
\end{cases}
\end{equation*}
Let $\mathcal{V}_e \subseteq \mathcal{V}$ be the set of actuated nodes that are controlled by an \emph{external} input and let $|\mathcal{V}_e|=p$.
We define
\be
e_{i}=\begin{cases}
1 \quad &i \in \mathcal{V}_e\\
0 \quad &\text{otherwise}.
\end{cases}
\ee
The dynamics of node $i \in \mathcal{V}$ are given by
\begin{subequations}
\begin{alignat}{2}
  T_{x_i}\dot x_i(t) =& -\sum_{k \in \mathcal{E}_i} B_{ik} \lambda_{k}(t) + e_{i} u_{i}(t) - d_i \\
  y_i(t) =&~h_i(x_i(t)),
\end{alignat}
\end{subequations}
where $x_i(t)$ is the storage (inventory) level, $u_{i}(t)$ the control input, $T_{x_i} \in \mathds{R}_{>0}$ a constant\footnote{Usually we have $T_{x_i} =1$ in the classical flow networks, where a material is transported. See however Subsection 7.2 for an example where $T_{x_i} \neq 1$.}, $d_i$ is a constant \emph{unknown} disturbance and $y_i = h_i(x_i)$ the measured output with $h_i(\cdot)$ a continuously differentiable and strictly increasing function. Moreover, $\mathcal{E}_i$ is the set of edges connected to node $i$ and $\lambda_{k}(t)$ is the flow on edge $k$.
We can represent the complete network compactly as\footnote{For the sake of simplicity,  the dependence of the variables on time $t$ is omitted in most of the remainder this paper.}
\begin{subequations}\label{model}
\begin{align}
T_x \dot x &= -B \lambda + E u - d \label{model_x}\\
y &=~h(x),
\end{align}
\end{subequations}
where $T_x \in \mathds{R}_{>0}^{n\times n}$,  $B \in \mathds{R}^{n\times m}$,  $\lambda \in \mathds{R}^m$,  $u \in \mathds{R}^p$ and $d \in \mathds{R}^n$. Without loss of generality we assume that only the first $p$ nodes have a controllable external input, {\it i.e.} $ \{1,\hdots, p \} = \mathcal{V}_{e}$, and consequently
$E \in \mathds{R}^{n\times p}$ is of the form
\be \label{def_of_E}
E=\begin{bmatrix}
      I_{p\times p} \\
      \mathbf{0}_{(n-p)\times p} \\
    \end{bmatrix}.
\ee
Furthermore, $y \in \mathds{R}^{n}$ and $h(x)\in \mathds{R}^{n}$ of which the $i$-th component is given by $h_i(x_i)$.
Throughout this work we will study the control of the inputs to the nodes and the control of the flows on the edges. We make two basic assumptions on the network that allows us to formulate the control objectives explicitly in the next section. First, in order to guarantee that each node can be reached from anywhere in the graph we make the following assumption on the topology:
\begin{assumption}[Connectedness]\label{connected_graph_dynamics}
The graph $\mathcal{G}$ is connected.
\end{assumption}
We recall (see {\it {\it e.g.}} \cite[Lemma 2.2]{bapat2010graphs}) the following useful lemma:
\begin{lemma}[Rank of $B$]\label{lemmaRank}
  Let $\mathcal{G}$ be a graph with $n$ nodes and let $B$ be the incidence matrix of $\mathcal{G}$. Then the rank of $B$ is $n-1$ if and only if $\mathcal{G}$ is connected.
\end{lemma}

Second, to compensate for the disturbances to the network, the following assumption is required:
\begin{assumption}[Controllable inputs]\label{minimal_one_input}
There is at least one node that has a controllable external input, {\it {\it i.e.}} $p\geq1$.
\end{assumption}
An immediate consequence of Assumption \ref{connected_graph_dynamics} and its related Lemma \ref{lemmaRank} is the following result:
\begin{lemma}[Rank of $
                                                                                                                   \begin{bmatrix}
                                                                                                                     B & E \\
                                                                                                                   \end{bmatrix}
                                                                                                                $]\label{lemmaBE}
If Assumption \ref{connected_graph_dynamics} is satisfied, then Assumption \ref{minimal_one_input} is equivalent to $
                                                                                                                   \begin{bmatrix}
                                                                                                                     B & E \\
                                                                                                                   \end{bmatrix}
                                                                                                                $ being full row rank, {\it i.e.} $\text{\rm rank}( \begin{bmatrix}
                                                                                                                     B & E \\
                                                                                                                   \end{bmatrix}) = n$.
\end{lemma}
Particularly, we will use the fact that the pseudoinverse of $ \begin{bmatrix}
                                                                                                                     B & E \\
                                                                                                                   \end{bmatrix}$ constitutes a right inverse, which has been exploited within a similar context in {\it e.g.} \cite{blanchini_2016_automatica}.

\section{Optimal regulation with input and flow constraints} \label{optimal_regulation}

In this section we discuss two control objectives and the various input and flow constraints under which the objectives should be reached.
We start with discussing the two objectives. The first objective is concerned with the output $y=h(x)$ in (\ref{model}), at steady state.
\begin{objective}[Output regulation] \label{objective1} Let $\overline y$ be a desired constant setpoint, then the output $y = h(x)$ of (\ref{model}) asymptotically converges to $\overline y$,  {\it i.e.}
  \begin{align}
\lim_{t \rightarrow \infty} \|h(x(t))-\overline {y}\|&= 0 . \label{xgoal}
\end{align}
\end{objective}

\begin{remark}[Tracking of a ramp]
In case that $h(x) = x$, Objective 1 can immediately be extended to the possibility of tracking a linear transition from the current setpoint $\overline y(t_1)$ to a new setpoint $\overline y(t_2)$ with $t_2 > t_1$.
To do so, the desired reference signal is modelled as a ramp, {\it i.e.}
\begin{align}
\overline y(t) =\overline y(t_1)  + \frac{t-t_1}{t_2 - t_1}(\overline y(t_2) - \overline y(t_1)).
\end{align}

  After a coordinate transformation $\tilde{x}(t)=x(t)-\overline y(t)$, we obtain a system of the same form as (\ref{model_x}), where the evolution of $\tilde x$ is described by
\be\label{eq:incremental}
T_x \dot{ \tilde{x}} = -B \lambda + E u -\tilde{d}.
\ee
The corresponding constant disturbance is now given by
\begin{align}
\tilde{d}=d + \frac{1}{t_2 - t_1}(\overline y(t_2) - \overline y(t_1)).
\end{align}
Note that boundedness of $\tilde{x}_i$ does not imply boundedness of $x_i$ as $\overline y_i(t)$  increases or decreases constantly over time. Therefore, the used invariance principle in the later sections is not immediately applicable if we consider the original variables of the system. Nevertheless, the subsequent analysis can be applied to the incremental system (\ref{eq:incremental}) if we consider $\tilde x$ as the state.  
\end{remark}
To ensure feasibility of Objective 1, the following assumption is made:
\begin{assumption}[Feasible setpoint]\label{assumesetpoint}
  The desired setpoint $\overline y$ satisfies 
  \begin{align}
    \overline y_i \in \mathcal{R}(h_i) \quad \text{ for all } i \in \mathcal{V}.
  \end{align}  
\end{assumption}
\stopmodif
At a state where $\overline x$ is constant and satisfies $h(\overline x)= \overline y$ system (\ref{model_x}) necessarily satisfies
 \begin{align}\label{ssequi}
\begin{split}
\boldsymbol{0}=&-B\lambda+ E u-d.\\
\end{split}
\end{align}
 Premultiplying (\ref{ssequi}) with $\mathds{1}_n^T$ results in
 \begin{align}
   0 = \mathds{1}_n^TE u - \mathds{1}_n^T d,
 \end{align}
 such that at a steady state the total input to the network needs to be equal to the total disturbance.
If there are two or more inputs  to the network ({\it {\it i.e.} $p\geq 2$}), it is natural to wonder if the total input can be  coordinated  optimally among the nodes.
To this end, we assign a strictly convex linear-quadratic cost function $C_i(u_{i})$ to each input of the form
\begin{align}
C_i(u_{i})=\frac{1}{2}q_i u_{i}^2 +  r_iu_{i}+ s_i, \label{individualcosts}
\end{align}
with $q_i\in\R_{>0}$ and $ r_i, s_i \in\R$. The total cost can be expressed as
\begin{align}\label{costs}
  \begin{split}
   C(u)=&~ \sum_{i \in \mathcal{V}_e} C_i(u_{i}) =  \frac{1}{2}u^T Q u + r^T u+ s,
  \end{split}
\end{align}
where $Q=\text{diag}(q_1, \dots, q_p )$, $r=(
                                r_1, \hdots, r_p
                            )^T$ and $s=\sum_{i \in \mathcal{V}_e} s_i$.
 Minimizing (\ref{costs}), while satisfying the equilibrium condition (\ref{ssequi}),
  gives rise to the following optimization problem:
\begin{equation}
\begin{aligned}\label{optimalprod}
& \underset{u, \lambda}{\text{minimize}}
& & C(u) \\
& \text{subject to}
& & \mathbf{0}=-B\lambda+E u-d.\\
\end{aligned}
\end{equation}
It is possible to explicitly characterize the solution to (\ref{optimalprod}) and we do so in the following lemma:
\begin{lemma}[Solution to optimization problem (\ref{optimalprod})] \label{lemma3} The solution to (\ref{optimalprod}) is given by
\be
\overline{u}  = Q^{-1} (\kappa - r),\label{qpopt}
\ee
where
\be \label{mu_opt}
  \kappa = E^T\frac{\mathds{1}_n\mathds{1}_n^T }{\mathds{1}_p^T Q^{-1}\mathds{1}_p}(d + E Q^{-1}r ).
\ee


\e1

\end{lemma}
\textbf{Proof.}
The proof follows standard arguments from convex optimization and from realizing (\cite[Lemma 4]{trip_2016_automatica}) that the constraint in (\ref{optimalprod}) can be equivalently replaced by
\begin{align}
  \mathds{1}_n^T (E u - d) = 0.
\end{align}
\qedp

\begin{remark}[Identical marginal costs]
Note that we can rewrite (\ref{qpopt}) as
\begin{align}
  \kappa =&~ Q\overline u + r,
\end{align}
and that $\kappa \in {\rm im}(\mathds{1}_p)$. It follows that, when evaluated at the solution to (\ref{optimalprod}),
the so-called marginal costs $\frac{\partial C_i(u_i)}{\partial u_i} = q_iu_{i} + r_i$ are identical
for all $i \in \mathcal{V}_e$ (\cite{hoy2011mathematics}).
\end{remark}
We are now ready to state the second control objective.

\begin{objective}[Optimal feedforward input] \label{objective2}
The input at the nodes asymptotically converge to the solution to (\ref{optimalprod}), {\it i.e.}
  \begin{align}
\lim_{t \rightarrow \infty} \|u(t)-\overline{u} \|=&~0, \label{u_d_goal}
\end{align}
with $\overline u$ as in (\ref{qpopt}).
\end{objective}

We now turn our attention to possible constraints on the control inputs $u$ and $\lambda$ under which the objectives should be reached.
First, in physical systems the input $u$ is generally constrained by a minimum value (often zero, preventing a negative input) and a maximum value, representing {\it e.g.} a production capacity.
\begin{constraint}[Input limitations]
The inputs at the nodes satisfy
  \begin{align}
u_i^-<&~ u_i(t) < u_i^+ \quad \text{for all } i\in\mathcal{V}_e \text{ and all } t \geq 0, \label{upsatbound}
\end{align}
with $u_i^-, u_i^+ \in \R$ being suitable constants.
\end{constraint}

Second, the
flows on the edges are often constrained to be
unidirectional and to be within the capacity of the edges.

\begin{constraint}[Flow capacity]
The flows on the edges satisfy
  \begin{align}
\lambda_k^- <&~ \lambda_k(t) < \lambda_k^+\quad \text{for all } k \in \mathcal{E} \text{ and all } t \geq 0,  \label{uesatbound}
\end{align}
with  $\lambda_k^-, \lambda_k^+ \in \R$ being suitable constants.
\end{constraint}
Note that physical limitations and safety requirements demand that the constraints should be satisfied \emph{for all time} and not only at steady state.
\begin{remark}[Special cases]
The unconstrained case can be regarded as a particular example of the considered setting. This is obtained by taking $-\infty$ as a lower and $\infty$ as an upper bound for both $u_{i}$ and $\lambda_{k}$.
Moreover, if we take $\lambda_k^- \geq 0$ or $\lambda_k^+ \leq 0$, the flow on edge $k$ is constrained to be unidirectional.
\end{remark}

In many applications it is desirable to have a distributed control architecture where controllers rely only on local information to decrease communications, to increase robustness and to improve the scalability of the control scheme. We therefore require that the controllers to be designed, only depend on information available from adjacent nodes in the physical flow network or adjacent nodes in a digital communication network that is deployed  to ensure optimality (see the next section).

For convenience, we summarize the objectives and constraints yielding the following controller design problem.
\begin{problem}[Controller design problem]
\label{problem1}
Design distributed controllers that regulate the external inputs $u$  at the nodes  and the flows $\lambda$ on the edges, such that
\begin{align}
\begin{split}
\lim_{t \rightarrow \infty} \|h(x(t))-\overline y\|&=0\\
\lim_{t \rightarrow \infty} \|u(t)-\overline{u} \|&=0,
\end{split}
\end{align}
 where  $\overline{y}$ is the desired setpoint and $\overline{u}$ is as in (\ref{qpopt}). Furthermore,
\begin{align}\label{constraints_problem1}
\begin{split}
\lambda_k^-<&~ \lambda_k(t) <\lambda_k^+\\
u_i^-<&~ u_i(t) <u_i^+,
\end{split}
\end{align}
 for all $k\in\mathcal{E}$, $i\in\mathcal{V}_e$ and $t\geq0$.
\end{problem}

\begin{remark}[Positive systems]
  A common requirement is that, additionally to Objective 1 and Objective 2, the state $x$ has to be nonnegative, {\it i.e.} $x(t) \geq \boldsymbol{0}$ for all $t$. Although, achieving output regulation, with $\overline x > \boldsymbol{0}$, is in practical cases sufficient to ensure that $x(t) \geq \boldsymbol{0}$ for all $t$, when the system  is suitably initialized (see also the case studies in Section \ref{case_study}), a theoretical guarantee is difficult to  obtain, due to the presence of an unknown and constant disturbance $d$. An interesting future endeavor is to study the design of controllers achieving Objective 1 and Objective 2 within the setting of so-called positive systems (\cite{benvenuti2002positive}, \cite{valcher_2014_tac}, \cite{arneson_2016_automatica}, \cite{ebihara_2017_tac}).
\end{remark}

\section{Controller design} \label{controller_design}

In this section we propose \emph{distributed} input and flow controllers that achieve the various objectives under the  constraints discussed in the previous section. The controllers will be designed to enjoy a passivity property and asymptotic stability of the closed loop system will derive from a suitable power preserving interconnection of the flow network and the controllers. Both the passivity property  as well as the stability of the closed loop system will be discussed in the next section.

Before introducing the controllers, we make two observation.
First, by premultiplying both sides of (\ref{model_x}) with $\mathds{1}_n^T$, we obtain that
\be \label{chap3:summed_dynamics}
\mathds{1}_n^T T_x \dot x =\mathds{1}_n^T ( E u - d),
\ee
which shows that the aggregated storage level are independent of the flows $\lambda$, that distribute the material \emph{within} the network.
Second, at steady state, (\ref{chap3:summed_dynamics}) becomes
\be \label{chap3:summed_dynamics_zero}
0 =\mathds{1}_n^T ( E \overline u - d),
\ee
which implies that a balance between the total input and disturbance is required to obtain a steady state. The first observation motivates the design of a flow controller, aiming at distributing the deviation from the desired output, $y - \overline y$, equally among the nodes, \emph{i.e.} $y_i - \overline y_i = y_j - \overline y_j$, for all $i,j \in \mathcal{V}$. The controllers at the nodes, regulating the external input to the network, are then designed to steer the deviation from the desired output to zero, by optimally allocating the external inputs to the network, such that the total input is identical to the total disturbance.  We start with discussing the flow controller in more detail.


\subsection{Flow controller}
\begin{figure}
\centering
  \includegraphics[trim=0cm 16cm 18cm 0cm, clip=true, width=0.40 \textwidth]{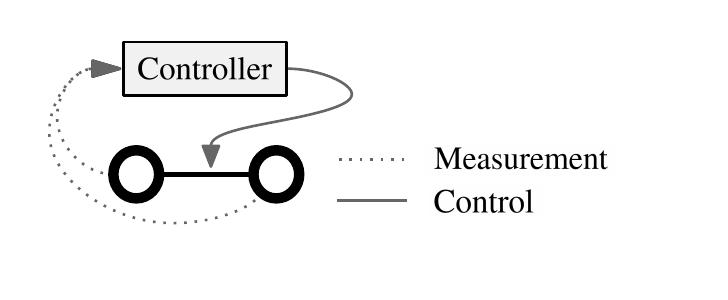}\\
  \caption{The controller that is located at the edge has access to the outputs of its adjacent nodes. Using these measurements as inputs, the controller generates the flow rate $\lambda$ on the edge.}\label{controller_of_link}
\end{figure}

We design a controller that regulates the flows on the edges, aiming at consensus in the error $y-\overline y$ (balancing), while obtaining a useful passivity property of the resulting closed loop system when interconnected with (\ref{model}). Consider the following controller:
\begin{align} \label{lambdadotpi}
\begin{split}
T_\mu \dot \mu =&  ~    B^T (h(x) - \overline y) - (f(\mu) - \xi) \\
T_\xi \dot \xi =&~ f(\mu)-\xi\\
\lambda=&~f(\mu),
  \end{split}
\end{align}
where $T_\mu, T_\xi \in \R^{m \times m}_{> 0}$  are diagonal matrices with strictly positive entries, $\mu, \xi \in \R^m$ and the mapping
$f(\cdot):\R^m \rightarrow \R^m$, with $f(\mu)=(f_1(\mu_1),\ldots, f_m(\mu_m))^T$,
 has suitable properties discussed in Assumptions \ref{assump_bounded_controller_outputs2} and \ref{assump_bounded_controller_outputs} below. Moreover, $B$ is the incidence matrix reflecting the topology of the physical network,
which implies that the flow controller on edge $k$ only requires information from its adjacent nodes (see also Figure \ref{controller_of_link}). Note that the term $[ B^T (h(x) - \overline y)]_k$ determines the difference in the output error of the two adjacent nodes to edge $k \in \mathcal{E}$. As will be discussed in Remark 9 and Subsection 6.2, the state $\xi$ is introduced to prove convergence to a constant flow, preventing oscillations. The passivity property, mentioned before, is derived in Lemma 6 in the next section.

\subsection{Controller at the nodes}
\begin{figure}
\centering
  \includegraphics[trim=0cm 14.7cm 14cm 0cm, clip=true, width=0.45 \textwidth]{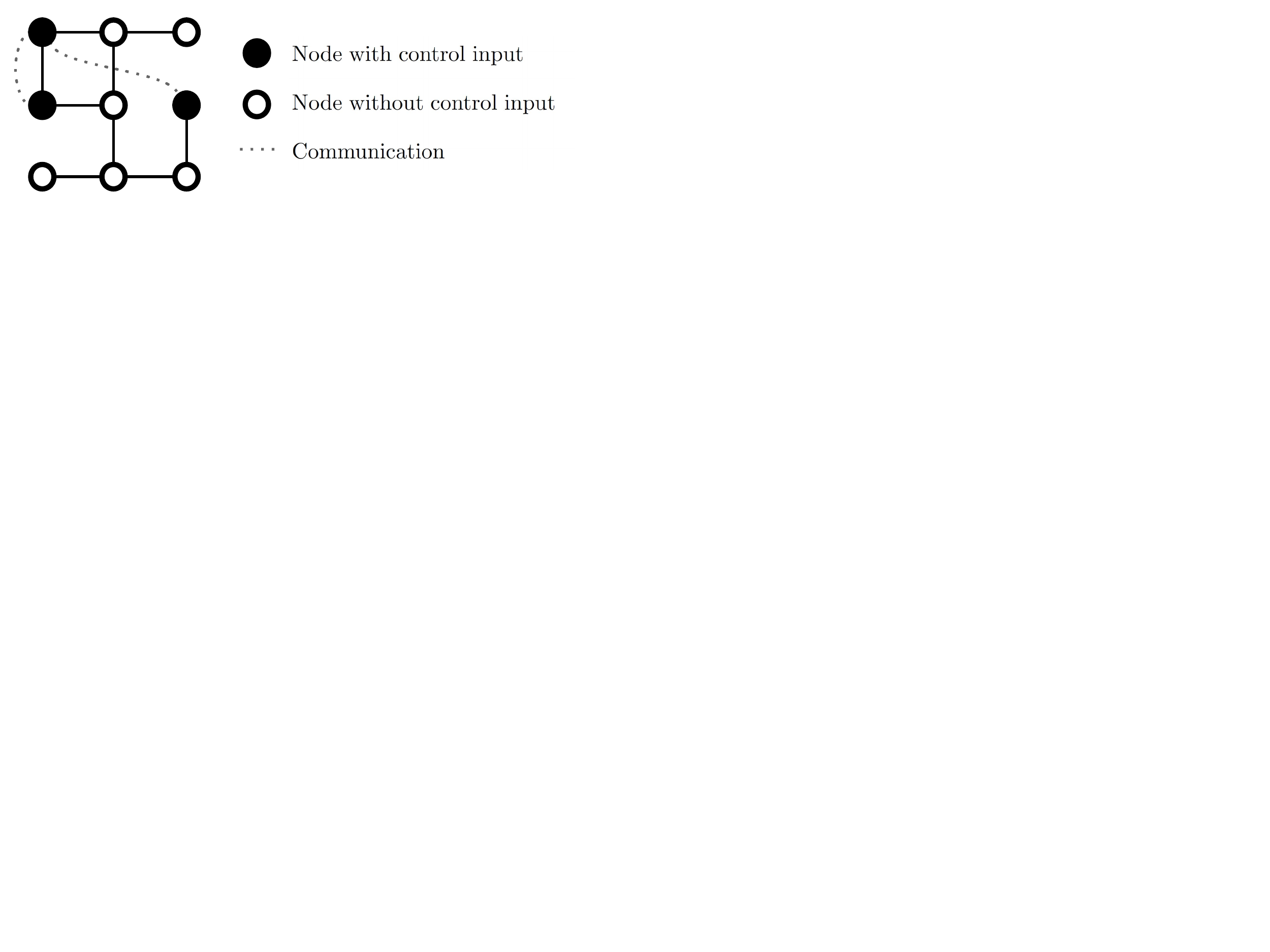}\\
  \caption{ Example of a flow network including a communication graph. }\label{communication_graph}
\end{figure}

Next, we design an input controller $u_i$ at each node $i$ that adjusts the external input to the network. Inspired by the result in \cite{trip_2017_tcns}, where a similar control problem is considered in the setting of power networks, we propose the controller
\begin{align} \label{controllerinput}
\begin{split}
T_\theta \dot \theta =&-E^T(h(x) - \overline y) -(g(\theta) - \phi)   \\
T_\phi \dot \phi =&~ g(\theta) -\phi - QL^{com}(Q \phi+r)\\
  u =&~ g(\theta),
  \end{split}
\end{align}
where $T_\theta, T_\phi \in \R_{>0}^{p \times p}$ are diagonal matrices with strictly positive entries, $\theta, \phi \in \R^p$ and the mapping
 $g(\cdot):\R^p \rightarrow \R^p$, with $g(\theta)=(g_1(\theta_1)\ldots g_p(\theta_p))^T$,
 has suitable properties discussed in Assumptions \ref{assump_bounded_controller_outputs2} and \ref{assump_bounded_controller_outputs} below.
Moreover, $L^{com}$ is the Laplacian matrix reflecting the communication topology (see also Figure \ref{communication_graph}). This communication ensures that, at steady state, a consensus is obtained in the marginal costs, {\it i.e.} $Qg(\overline \theta) + r \in \text{im}(\mathds{1}_p)$. In order to guarantee that all marginal costs converge to the same value we make the following assumption on the communication network.

\begin{assumption} [Communication network]\label{assump:com}
  The graph reflecting the communication topology is balanced\footnote{A directed graph is balanced if the (weighted) in-degree is equal to the (weighted) out-degree of every node.  } and strongly connected.
\end{assumption}
\begin{lemma}[Consequence of Assumption \ref{assump:com}]\label{lemma:Lcomkernel}
If Assumption \ref{assump:com} is satisfied, then $L^{com}$ is a positive semi-definite matrix and
\be \label{Lcomkernel}
\phi^TL^{com}\phi=0,
\ee
if and only if $\phi \in{\rm im}(\mathds{1}_p)$.
\end{lemma}
\textbf{Proof.} The proof follows immediately from \cite[Theorem 7]{olfati2004consensus}. Specifically, since the communication graph is balanced, $L^{com}$ is positive semi-definite and (\ref{Lcomkernel}) satisfies
\be
\phi^TL^{com}\phi=\frac{1}{2}\phi^T(L^{com}+(L^{com})^T)\phi= \phi^T \hat{L}^{com} \phi,
\ee
where $\hat{L}^{com}$ is a Laplacian matrix  corresponding to the communication network with \emph{undirected} edges. Furthermore, ${\rm ker}(\hat{L}^{com}) = {\rm im}(\mathds{1}_p)$, due to the connectendness of the communication network.
 \qedp


%
Again, we introduced an additional state $\phi$, to ensure convergence to a constant point, whereas the term $[E^T(h(x) - \overline y)]_i$ provides an integral action to reduce the output error at the node $i \in \mathcal{V}_e$.

\begin{remark}[Local and exchanged information] According to (\ref{controllerinput}), every controller at node $i \in \mathcal{V}_e$, measures $y_i = h_i(x_i)$ and compares it with the desired set point $\overline y_i$. Information on the  marginal costs ($q_i \phi_i + r_i$) is exchanged among neighbours over a communication network with a topology described by $L^{com}$. Controller (\ref{controllerinput}) is therefore fully distributed. The output $g_i(\theta_i)$ is chosen to satisfy Constraint 1, and is discussed in more detail in the next subsection.
\end{remark}

\subsection{Feasibility of the control problem}


To ensure feasibility of the controller design problem, we impose two assumptions on the controllers (\ref{lambdadotpi}) and (\ref{controllerinput}).
The first assumption guarantees that the controllers are able to generate a (feedforward) control signal, that is required to attain a steady state of the system.
\begin{assumption}[Attainability of the steady state] \label{assump_bounded_controller_outputs2}
 ~Consider functions $f_k(\mu_k)$ and $g_i(\theta_i)$, in respectively (\ref{lambdadotpi}) and (\ref{controllerinput}).
 Let $\overline u$ be as in (\ref{qpopt}).  There exists\footnote{If  $B \overline \lambda=  E \overline u-d$ has any solution $\overline \lambda$, then all solutions are given by $ \overline \lambda = B^\dagger (E\overline u -d) +(I-B^\dagger B) \omega$, for an arbitrary vector $\omega \in \mathds{R}^m$, where $B^\dagger$ denotes the Moore-Penrose pseudoinverse of $B$. The existence of a solution $\overline \lambda$ is shown in the proof of Lemma \ref{lem:steady_state}.} a $\omega\in\mathds{R}^m$, such that $[B^\dagger (E\overline u -d) +(I-B^\dagger B) \omega]_k \in \mathcal{R}(f_k)$  for all $k\in\mathcal{E}$. Furthermore, $\overline u_i\in \mathcal{R}(g_i)$ for all $i\in\mathcal{V}_e$.
\end{assumption}
Moreover,  the controllers (\ref{lambdadotpi}) and (\ref{controllerinput}) can be designed to satisfy constraints (\ref{upsatbound}) and (\ref{uesatbound}),  by properly selecting $f(\mu)$ and $g(\theta)$.
Since $\lambda = f(\mu)$ and $u = g(\theta)$, the following assumption is sufficient to ensure that the inputs and flows do not exceed their limitations.
\begin{assumption}[Controller outputs] \label{assump_bounded_controller_outputs}
Functions $f_k(\cdot)$ and $g_i(\cdot)$, in respectively (\ref{lambdadotpi}) and (\ref{controllerinput}), are continuously differentiable, strictly increasing and satisfy
\begin{align}
\begin{split}
\mathcal{R}(f_k) &= (\lambda_k^-,\lambda_k^+) \label{chap3:uf+limit}\\
\mathcal{R}(g_i) &= (u_i^-,u_i^+),
 \end{split}
\end{align}
 for all $k\in\mathcal{E}$ and all $i\in\mathcal{V}_e$.
\end{assumption}
\stopmodif
The property of  $f_k(\mu_k)$ and $g_i(\theta_i)$  being continuously differentiable and strictly increasing functions, is exploited within the various proofs to establish the global convergence properties, and ensures \emph{e.g.} the existence of an inverse function.
Possible choices for $f_k(\mu_k)$ and $g_i(\theta_i)$, that satisfy Assumption \ref{assump_bounded_controller_outputs}, include {\it e.g.} the function $\gamma(z) = z$ in absence of any constraints, and also, upon proper scaling, the constraint enforcing functions $\gamma(z) = \tanh (z)$, $\gamma(z) = \arctan (z)$  (see also the case studies in Section \ref{case_study}).   \stopmodif

Before we analyse the stability of the system we investigate the properties of the steady state. To do so, we write system (\ref{model}) in closed loop with controllers (\ref{lambdadotpi}) and (\ref{controllerinput}), obtaining
\begin{subequations}\label{ode_complete}
 \begin{align}\label{ode_complete1}
\begin{split}
T_x \dot x =&
-B f(\mu) + E g(\theta) - d\\
T_\mu  \dot \mu =&  ~    B^T (h(x) - \overline y) -   ( f(\mu)- \xi)\\
T_\xi  \dot \xi =&  ~   f(\mu)- \xi
\end{split}\\
\begin{split} \label{ode_complete2}
T_\theta \dot \theta =&  -E^T(h(x) - \overline y) -(g(\theta) - \phi) \\
T_\phi \dot \phi =&~ g(\theta)-\phi - QL^{com}(Q \phi+r).
  \end{split}
\end{align}
\end{subequations}
Any equilibrium of system (\ref{ode_complete}) satisfies
\begin{subequations}\label{equilibrium}
\begin{align}
 \mathbf{0}  &= - B f(\overline \mu)  + E g(\overline \theta) -d       \label{uno} \\
 \mathbf{0}   &= B^T (h(\overline x) - \overline y)- (f(\overline \mu)-\overline \xi) \label{due} \\ 
 \mathbf{0}    &= f(\overline \mu)-\overline \xi \label{tre}\\
 \mathbf{0}    &= -E^T (h(\overline x) - \overline y) - (g(\overline \theta)-\overline \phi) \label{quattro}\\
 \mathbf{0}    &= (g(\overline \theta)-\overline \phi) - Q L^{com} (Q\overline \phi+r)\label{cinque}.
 \end{align}
\end{subequations}
We will now show that under  Assumptions \ref{connected_graph_dynamics}--\ref{assump_bounded_controller_outputs} there exists at least one solution to (\ref{equilibrium}) and all solutions (\ref{equilibrium}) satisfy the control objectives.


\begin{lemma}[Equilibria]\label{lem:steady_state}
Let Assumptions \ref{connected_graph_dynamics}--\ref{assump_bounded_controller_outputs2} hold. Then, there exists an equilibrium $(\overline x, \overline \mu, \overline \xi, \overline \theta, \overline \phi)$ of system (\ref{ode_complete}). Moreover, any equilibrium is such that $h(\overline x) = \overline y$ and
$g(\overline \theta) = \overline u$, where $\overline u$ is the optimal control input given by (\ref{qpopt}).
\end{lemma}
\textbf{Proof.}
 To prove the statement, we first show that at least one equilibrium of system (\ref{ode_complete}) exists.
 By Assumption \ref{assump_bounded_controller_outputs2}, $\overline u \in \mathcal{R}(g)$, and we set $\overline \theta = g^{-1}(\overline u)$.
  Also, we set $\overline \phi = \overline u$. Bearing in mind that $Q\overline u +r \in {\rm im}(\mathds{1}_p)$, we have that (\ref{cinque}) holds.
Furthermore, by definition, $\overline u$ satisfies
$
\mathds{1}^T_n(E \overline u - d)=0.
$
 Since the graph is connected (Assumption \ref{connected_graph_dynamics}) and ${\rm im}(B) = ({\rm ker}(B^T))^\perp =  ({\rm im}(\mathds{1}_n))^\perp $, we have that $E \overline u - d\in{\rm im}(B)$. For this reason, there exists a $\overline \lambda$ satisfying
$-B\overline \lambda +E \overline u - d =\mathbf{0}$, and any solution is given by
$
\overline\lambda=B^\dagger (E\overline u -d) +(I-B^\dagger B) \omega,
$
for an arbitrary vector $\omega\in\mathds{R}^m$. By Assumption  \ref{assump_bounded_controller_outputs2}, there exists at least one $\omega$ such that $\overline\lambda \in \mathcal{R}(f_k)$. Taking such a $\overline \lambda$,
setting $\overline \xi = \overline \lambda$ and $\overline \mu= f^{-1}(\overline \lambda)$, shows that (\ref{uno}), (\ref{tre}) hold. Since $\overline y \in \mathcal{R}(h)$ (Assumption \ref{assumesetpoint}), setting $\overline x = h^{-1}(\overline y)$ shows (\ref{due}) and (\ref{quattro}). Hence, there exists a state $(\overline x, \overline \mu, \overline \xi, \overline \theta, \overline \phi)$ that satisfies the equations (\ref{equilibrium}) and is therefore an equilibrium of (\ref{ode_complete}). \stopmodif

 Next, we show that any equilibrium $(\overline x, \overline \mu, \overline \xi, \overline \theta, \overline \phi)$ necessarily satisfies $h(\overline x) = \overline y$ and
$g(\overline \theta) = \overline u$, where $\overline u$ is the optimal control input given by (\ref{qpopt}).
From \eqref{tre}, $\overline \xi=f(\overline \mu)$ holds and we will show that this implies that necessarily $h(\overline x) = \overline y$.
By \eqref{cinque}, bearing in mind that $L^{com}$ is the Laplacian of a balanced and strongly connected graph (Assumption \ref{assump:com}), we have according to Lemma \ref{lemma:Lcomkernel}  that $\mathds{1}_{p} ^T Q^{-1} (g(\overline \theta)-\overline \phi)=0$. This, together with  \eqref{quattro}, implies that
$\mathds{1}_{p} ^T Q^{-1}  E^T (h(\overline x) - \overline y)= 0$. By \eqref{due} and $\overline \xi=f(\overline \mu)$, we also have $B^T (h(\overline x) - \overline y)= \mathbf{0}$. Hence,
\begin{align}
\begin{bmatrix}
    \mathds{1}_{p} ^T Q^{-1}  E^T \\
    B^T \\
  \end{bmatrix}
(h(\overline x) - \overline y)=\mathbf{0}.
\end{align}
We now prove that necessarily $h(\overline x) - \overline y=\mathbf{0}$.
Suppose, \emph{ad absurdum}, that there exists $v\ne \mathbf{0}$ such that
\begin{align}
\begin{bmatrix}
    \mathds{1}_{p} ^T Q^{-1}  E^T \\
    B^T \\
  \end{bmatrix}v=\mathbf{0}.
\end{align}
 By Assumption \ref{connected_graph_dynamics}, it follows that $v=\mathds{1}_{n} v_*$ with $v_*$ a scalar.  Then $\mathds{1}_{p} ^T Q^{-1}  E^T \mathds{1}_{n} v_*=0$, which is, by definition of $E$ in (\ref{def_of_E}) and Assumption \ref{minimal_one_input}, equivalent to $\mathds{1}_{p} ^T Q^{-1}  \mathds{1}_{p} v_*=0$. This implies that $v_*=0$, contradicting that $v = \mathds{1}_n v^* \neq \boldsymbol{0}$. Hence, necessarily $h(\overline x) - \overline y=\mathbf{0}$ and by strict monotonicity of $h(\cdot)$, we must have that $\overline x= h^{-1}(\overline y)$.

Since $h(x) = \overline y$, it follows from (\ref{quattro}) that $ g(\overline \theta)=\overline\phi$, and by strict monotonicity of $g(\theta)$, that $ \overline \theta=g^{-1}(\overline\phi)$. Moreover, from (\ref{cinque}) we obtain that $L^{com} (Q\overline \phi+r)=\boldsymbol{0}$, and since the communication graph is strongly connected due to Assumption \ref{assump:com}, we have that $Q\overline \phi+r \in {\rm im}(\mathds{1}_p)$.  Since $\mathds{1}_{n}^TB=\boldsymbol{0}$, we obtain from (\ref{uno}) that $\mathds{1}_{n}^T(Eg(\overline\theta)-d) = 0$. Bearing in mind that $\overline u$ satisfies $Q\overline u + r \in \text{\rm im}(\mathds{1}_p)$ and $\mathds{1}_n^T(E\overline u -d) = 0$,
we have consequently that $g(\overline \theta)=\overline\phi = \overline u$, with $\overline u$ as in (\ref{qpopt}).
\stopmodif
\qedp

As a consequence of Lemma \ref{lem:steady_state} we have that if Assumptions \ref{connected_graph_dynamics}--\ref{assump_bounded_controller_outputs2} hold, system (\ref{ode_complete}) is equivalent to
\begin{subequations}\label{ode_complete_inc}
 \begin{align}\label{ode_complete_inc1}
\begin{split}
T_x \dot x =&
-B (f(\mu)-f(\overline \mu)) + E (g(\theta)-g(\overline \theta))\\
T_\mu  \dot \mu =&    ~    B^T (h(x) - h(\overline x)) \\
&- ((f(\mu)-f(\overline \mu))  - (\xi-\overline\xi))\\
T_\xi  \dot \xi =&  ~  (f(\mu)-f(\overline \mu))-(\xi-\overline\xi)
\end{split}\\
\begin{split}\label{ode_complete_inc2}
T_\theta \dot \theta =&  -E^T(h(x) - h(\overline x))\\
& -(g(\theta)-g(\overline \theta)) + (\phi-\overline \phi) \\
T_\phi \dot \phi =& ~(g(\theta)-g(\overline \theta))-(\phi-\overline \phi) \\
&- QL^{com}Q(\phi-\overline \phi),
  \end{split}
\end{align}
\end{subequations}
a form that will be exploited in the stability analysis.

\section{Stability analysis} \label{stability_analysis}
In this section we analyze the stability of the closed-loop system (\ref{ode_complete}). The analysis is foremost based on LaSalle's invariance principle and exploits useful properties of interconnected incrementally passive systems. To facilitate the discussion, we first recall the following definition:

\begin{definition}[Incremental passivity]
  System \begin{align} \begin{split}
  \dot x =& ~f(x,u)\\
   y=&~h(x),
  \end{split}
  \end{align}
  $x \in \mathcal{X}$, $\mathcal{X}$ the state space, $u,y \in \mathds{R}^n$, is incrementally passive\footnote{With some abuse of terminology, we state the incremental passivity property with respect to a steady state solution. This is in contrast to the `usual' definition where the incremental passivity property holds with respect to any solution (\cite{Pavlov2008}).} with respect to a constant triplet $(\overline x, \overline u, \overline y)$ satisfying
  \begin{align}
  \begin{split}
    \boldsymbol{0} =& ~f(\overline x, \overline u) \\
    \overline y =& ~h(\overline x),
    \end{split}
  \end{align}
  if there exists a continuously differentiable and  radially unbounded  function $V(x, \overline x) : \mathcal{X} \rightarrow \mathds{R}$, such that for all $x \in \mathcal{X}$, $u \in \mathds{R}^m$ and $y=h(x)$, $\overline y = h(\overline x)$
  \be
  \dot{V}(\cdot) = \frac{\partial V}{\partial x} f(x,u) + \frac{\partial V}{\partial \overline x} f(\overline x,\overline u) \leq (y-\overline y)^T (u-\overline u).
\ee
\end{definition}

We now proceed with establishing the incremental passivity property of (\ref{ode_complete1}), that is the proposed flow controller (\ref{lambdadotpi}) renders the network dynamics (\ref{model}) incrementally passive with respect to the input $ E g(\theta)$ and output $h(x)$.
\begin{lemma}[Incremental passivity of (\ref{ode_complete1})]\label{incr1}
Let Assumptions \ref{connected_graph_dynamics}--\ref{assump_bounded_controller_outputs2} hold. System
(\ref{ode_complete1}) with input $Eg(\theta)$ and output $h(x)$ is incrementally passive with respect to the constant  $(\overline x, \overline \mu, \overline\xi)$ satisfying
(\ref{uno})-(\ref{tre}).
Namely, the radially unbounded storage function
$V_1(x, \overline x, \mu, \overline \mu, \xi, \overline \xi)$ satisfies
\be
\begin{aligned} \label{v1dot}
\dot V_1(\cdot)  =& ~(h(x) - h(\overline x))^T E(g(\theta) - g(\overline \theta))\\
&-(f(\mu)-\xi)^T(f(\mu)-\xi),
\end{aligned}
\ee
along the solutions to (\ref{ode_complete1}).
\end{lemma}
\textbf{Proof.}
Consider the storage function
\begin{align}\label{v1}
\begin{split}
  V_1(x, \overline x, \mu, \overline \mu, \xi, \overline \xi) =&
  \sum_{i \in \mathcal{V}} T_{x_i} \int_{\overline x_i}^{x_i} h_i(y) -h_i(\overline x_i)dy\\
   &+\sum_{k \in \mathcal{E}} T_{\mu_k} \int_{\overline \mu_k}^{\mu_k} f_k(y) -f_k(\overline \mu_k)dy\\
  &+ \frac{1}{2}(\xi - \overline \xi)^T T_\xi (\xi - \overline \xi).
\end{split}
\end{align}
Since $h_i(x_i)$ and $f_k(\mu_k)$ are strictly increasing functions, the incremental storage function $V_1(\cdot)$ is radially unbounded.
Furthermore, $V_1(\cdot)$ satisfies along the solutions to (\ref{ode_complete1}), or equivalently along the solutions to (\ref{ode_complete_inc1}),
 \begin{align}
 \begin{split}
  \dot{V}_1(\cdot) =& ~(h(x) - h(\overline x))^T T_x \dot{x} + (\xi-\overline \xi )^T T_\xi \dot{\xi}\\
   & \quad + (f(\mu) -f(\overline \mu))^T T_\mu \dot\mu \\
  =& ~(h(x) - h(\overline x))^T    E (g(\theta) - g(\overline \theta))\\
  &-((f(\mu)-f(\overline\mu))-(\xi-\overline \xi))^T\\
  &\quad \cdot((f(\mu)-f(\overline\mu))-(\xi-\overline \xi)),
    \end{split}
\end{align}
 Since $f(\overline\mu) = \overline \xi$, $V_1(\cdot)$ indeed satisfies $(\ref{v1dot})$  along the solutions to (\ref{ode_complete1}).
\qedp

We now prove a similar result for (\ref{ode_complete2}), that is the controller (\ref{controllerinput}) is incrementally passive with respect to the input $-h(x)$ and output $Eg(\theta)$.
\begin{lemma}[Incremental passivity of (\ref{ode_complete2})]\label{incr2}
Let Assumptions \ref{connected_graph_dynamics}--\ref{assump_bounded_controller_outputs2} hold. System
(\ref{ode_complete2}) with input  $-h(x)$ and output $Eg(\theta)$
is incrementally passive with respect to $(\overline \theta, \overline \phi)$  satisfying (\ref{quattro})-(\ref{cinque}).
Namely, the radially unbounded storage function
$V_2(\theta,  \overline \theta, \phi, \overline \phi)$ satisfies
\begin{align}\label{V2dot}
\begin{split}
\dot V_2(\cdot)  =& -(g(\theta) - \phi)^T(g(\theta) - \phi)\\
& -  (\phi -\overline \phi)^TQL^{com}Q (\phi -\overline \phi) \\
& - (g(\theta) - g(\overline \theta))^T E^T(h(x) - h(\overline x)),
\end{split}
\end{align}
along the solutions to (\ref{ode_complete2}).
\end{lemma}

\textbf{Proof.}
Consider the storage function
\begin{align}\label{v2}
\begin{split}
  V_2(\theta,  \overline \theta, \phi, \overline \phi) &= \sum_{i \in \mathcal{V}} T_{\theta_i} \int_{\overline \theta_i}^{\theta_i} g_i(y) -g_i(\overline \theta_i)dy\\
   &\quad + \frac{1}{2}(\phi - \overline \phi)^TT_\phi(\phi - \overline \phi).
   \end{split}
   \end{align}
  Note that since $g_i(\theta_i)$ is a strictly increasing function, the incremental storage function $V_2(\cdot)$ is radially unbounded.
Furthermore, $V_2(\cdot)$ satisfies along the solutions to (\ref{ode_complete2}), or equivalently along the solutions to (\ref{ode_complete_inc2}),
 \begin{align}
 \begin{split}
  \dot{V}_2(\cdot) =&~ (g(\theta) - g(\overline \theta))^T T_\theta \dot{\theta} \\
   & + (\phi -\overline \phi)^T T_\phi \dot\phi \\
=& ~(g(\theta) - g(\overline \theta))^T \cdot\left(-(g(\theta)-g(\overline\theta)) \right. \\
&   \left. + (\phi-\overline\phi) - E^T(h(x) - \overline y)   \right)  \\
   &  + (\phi -\overline \phi)^T (-(\phi-\overline \phi) + (g(\theta)-g(\overline \theta)) \\
   &- QL^{com}Q (\phi-\overline \phi)) \\
=&- (g(\theta) - g(\overline \theta))^T (g(\theta)-g(\overline\theta))\\
& + 2(g(\theta) - g(\overline \theta))^T (\phi-\overline\phi)  \\
 &  - (\phi -\overline \phi)^T (\phi-\overline \phi)\\
&- (\phi -\overline \phi)^TQ L^{com}Q (\phi-\overline \phi)\\
& - (g(\theta) - g(\overline \theta))^T E^T(h(x) - \overline y),    \\
        \end{split}
\end{align}
Since $g(\overline \theta) = \overline \phi$, $V_2(\cdot)$ indeed satisfies (\ref{V2dot}) along the solutions to (\ref{ode_complete2}).
\qedp

Exploiting the previous lemmas, we are now ready to prove the main result of this paper.
\begin{theorem}[Solving Problem 1 for system (\ref{model})]\label{theorem1}
  Let Assumptions \ref{connected_graph_dynamics}--\ref{assump_bounded_controller_outputs} hold. The solutions to system (\ref{model}), in closed loop with (\ref{lambdadotpi}) and (\ref{controllerinput}), globally converge to a point in the set
 \be
 \label{setUpsilon_1}
\Upsilon_1=\left\{ x, \mu, \xi, \theta, \phi \left| \begin{tabular}{l}
                                                      $B(f(\mu)-f(\overline \mu))=\boldsymbol{0}$,  \\
                                                      $B(\xi-\overline \xi)=\boldsymbol{0}$,  \\
                                                      $x=\overline x$, $\theta=\overline \theta$, $\phi=\overline \phi$\\
                                                     \end{tabular} \right.\right\},
\ee
where  $\lambda = f(\mu)$  is a constant, $h(x) = \overline y$ and where  $u = g(\theta)  = \overline u$, with $\overline u$ the optimal input given by (\ref{qpopt}).
Moreover, $u=g(\theta)$ and $\lambda=f(\mu)$ satisfy constraints (\ref{upsatbound}) and (\ref{uesatbound}) $\text{for all } t \geq 0$. Therefore, controllers (\ref{lambdadotpi}) and (\ref{controllerinput}) solve Problem \ref{problem1} for the flow network (\ref{model}).
\end{theorem}
\textbf{Proof.}
Satisfying constraints (\ref{upsatbound}) and (\ref{uesatbound}) $\text{for all } t \geq 0$ follows from the design of $g(\theta)$ and $f(\mu)$ and Assumption \ref{assump_bounded_controller_outputs}.
Let
\begin{align}\label{storagefunction}
V(\cdot)=V_1(x, \overline x, \mu, \overline \mu, \xi, \overline \xi)+V_2(\theta, \overline \theta, \phi, \overline \phi),
 \end{align}
 with $V_1(\cdot)$ and $V_2(\cdot)$ given in Lemma \ref{incr1} and Lemma \ref{incr2}, respectively. Consequently, $V(\cdot)$ satisfies
\be \label{lemma_v_dot}
\begin{aligned}
\dot{V}(\cdot) =&  -  (\phi -\overline \phi)^T  Q L^{com}Q (\phi-\overline \phi) \\
   & ~ - (g(\theta) - \phi)^T (g(\theta) - \phi) \\
   &~ -(f(\mu)-\xi)^T(f(\mu)-\xi),
\end{aligned}
\ee
 along the solutions to (\ref{ode_complete}). From (\ref{lemma_v_dot}) and Lemma \ref{lemma:Lcomkernel} we have that $\dot V ({\cdot}) \leq 0$, and since $V(\cdot)$ is radially unbounded, the solutions to (\ref{ode_complete}) approach the largest invariant set contained entirely in the set $\mathcal{S}_1$, where $\dot V (\cdot)= 0$. This set is characterized by
\be \label{setS1}
\mathcal{S}_1=\left\{ x, \mu, \xi, \theta, \phi \left| \begin{tabular}{l}
                                                      $\phi = g(\theta)$, $\xi = f(\mu)$, \\
                                                      $Q(\phi -\overline \phi)  \in \text{\rm im}(\mathds{1})$ \\
                                                     \end{tabular} \right.\right\},
\ee
where $Q(\phi + \overline \phi)  \in \text{im}(\mathds{1}_p)$  follows from Lemma \ref{lemma:Lcomkernel}.
On the set $\mathcal{S}_1$, system (\ref{ode_complete}) therefore satisfies
\begin{subequations}\label{invariantset1}
 \begin{align}
T_x \dot x =&
-B (f(\mu)-f(\overline \mu)) +   E(\phi- \overline \phi)  \label{invariantset1x} \\
T_\mu  \dot \mu =&    ~   B^T (h(x) - h(\overline x)) \label{invariantset1mu} \\
T_\xi  \dot \xi =&  ~  \boldsymbol{0} \label{invariantset1xi}\\
T_\theta \dot \theta =&  -E^T(h(x) - h(\overline x))\label{invariantset1theta}\\
T_\phi \dot \phi =& ~ \boldsymbol{0}. \label{invariantset1phi}
\end{align}
\end{subequations}
Due to (\ref{setS1}),  (\ref{invariantset1xi}) and (\ref{invariantset1phi}) we have that
 \begin{align}
\dot \mu =&~\left(\frac{\partial f(\mu)}{\partial \mu}\right)^{-1} \dot  \xi = \boldsymbol{0}\label{eqdotmu}\\
\dot \theta =&~\left(\frac{\partial g(\theta)}{\partial \theta}\right)^{-1} \dot \phi = \boldsymbol{0}. \label{eqdottheta}
\end{align}
where we note that $\frac{\partial f(\mu)}{\partial \mu}\neq\boldsymbol{0}$ and  $\frac{\partial g(\theta)}{\partial \theta}\neq\boldsymbol{0}$. It follows now from (\ref{invariantset1mu}), (\ref{invariantset1theta}), (\ref{eqdotmu}) and (\ref{eqdottheta}) that
\be
\begin{bmatrix}\label{parastart}
   ~~ B^T \\
   - E^T \\
\end{bmatrix}(h(x) - h(\overline x))=\boldsymbol{0}.
\ee
From Lemma \ref{lemmaBE} we recall that $\begin{bmatrix}
   B &
   - E
\end{bmatrix}^T$ has full column rank and therefore has a left inverse. As a result, we have that necessarily $h(x) - h(\overline x)=\boldsymbol{0},$ {\it i.e.} $h(x) = \overline y$. By strict monotonicity of $h(x)$, it follows that on the invariant set $x = \overline x$ and that $\dot x = \boldsymbol{0}$.

Premultiplying both sides of (\ref{invariantset1x}) %
by $\mathds{1}_p^T$, yields $0 = \mathds{1}_p^T(\phi- \overline \phi)$ and since $Q(\phi -\overline \phi)  \in \text{im}(\mathds{1}_p)$, where $Q$ is a diagonal matrix with only strictly positive entries, it follows that on the set where $\dot{V}=0$ necessarily $\phi= \overline \phi$.
From (\ref{setS1}) and (\ref{invariantset1x})  it therefore follows that $B(f(\mu)-f(\overline \mu))=\boldsymbol{0}$ and                                                     $B(\xi-\overline \xi)=\boldsymbol{0}$. Moreover, since on the set  $\mathcal{S}_1$, $\phi=g(\theta)$ and  $\overline \phi=g(\overline\theta)$, we also have that $g(\theta)= g(\overline \theta)=\overline u$ (see also Lemma \ref{lem:steady_state}).
Consequently, system (\ref{ode_complete}) indeed approaches the set $\Upsilon_1$, where $h(x) =\overline y$ and where
$u = g(\theta) = g(\overline \theta) = \overline u$,   with $\overline u$ the optimal input given by (\ref{qpopt}).
To prove convergence to a point in the set $\Upsilon_1$, we note that $\Upsilon_1$ consists of equilibria of (\ref{ode_complete}).
Since the incremental storage function $V(\cdot)$ can be defined with respect to any equilibrium in $\Upsilon_1$, and since $V(\cdot) \leq 0$,  every point in $\Upsilon_1$ is a  Lyapunov stable equilibrium of system (\ref{ode_complete}). Consequently, every positive limit set associated with any solution to system (\ref{ode_complete}) consists of Lyapunov stable equilibria. It then follows by \cite[Theorem 4.20]{haddad2008nonlinear} that this positive limit set is a singleton, which proves convergence to a point. 
\qedp

\begin{remark}[Uniqueness of $\mu$]
In the case that the graph $\mathcal{G}$ contains no cycles, {\it {\it i.e.} } the graph is a tree, then there exists a unique solution $\mu$, to $B(f(\mu)-f(\overline\mu))=\boldsymbol{0}$.
\end{remark}

\begin{remark}[Locally increasing mappings]
  Note that the global convergence result is a consequence of the strictly increasing behavior of the nonlinear functions $f_k(\mu_k)$, $g_i(\theta_i)$ and $h_i(x_i)$. In case the functions are increasing on a finite interval, a local result of Theorem 1 can be derived.  An important class of functions for which this holds are functions that are not necessary increasing on the whole domain, such as sinusoidal functions.
\end{remark}

\begin{remark}[Avoiding oscillations] \label{rem:avoiding_oscillation} In the proof of Theorem \ref{theorem1}, we exploited the dynamics of the additional control variables $\xi$ and $\phi$ to conclude that on the invariant set $\dot \mu = \dot \theta = \boldsymbol{0}$. It is natural to wonder if these additional controller states are essential to obtain the convergence result of Theorem \ref{theorem1}. Therefore, we compare (\ref{lambdadotpi}) and (\ref{controllerinput}) with controllers of the form
\begin{align}
\begin{split}\label{alto2}
T_\mu \dot \mu =&  ~B^T (h(x) - \overline y) \\
  \lambda =&~ f(\mu)
 \end{split}\\
\begin{split}\label{alto1}
 T_\theta \dot \theta =& - QL^{com}(Q g(\theta)+r)  - (h(x) - \overline y)\\
  u =&~ g(\theta),
 \end{split}
\end{align}
as both (\ref{lambdadotpi})-(\ref{controllerinput}) and (\ref{alto2})-(\ref{alto1}) admit a steady state where $h(\overline x) = \overline y$ and $g(\overline \theta)= \overline u$. However, in contrast to (\ref{ode_complete}), for which we have proven global convergence to the desired state, system
\begin{align}\label{alto_complete}
\begin{split}
T_x \dot x =& -B f(\mu) +E g(\theta) - d\\
T_\mu \dot \mu =&  ~B^T (h(x) - \overline y) \\
T_\theta \dot \theta =& - QL^{com}(Q g(\theta)+r)-(h(x) - \overline y),
  \end{split}
\end{align}
can converge (depending on $E$ and $Q$) to a limit cycle exhibiting oscillatory behavior as has been shown in \cite{scholten2016optimal}.
To illustrate this claim, consider the linear case, where $f(\mu)=\mu $, $g(\theta)=\theta$ and $h(x)=x$. 
Introducing $\tilde x = x - \overline x$, $\tilde  \mu = \mu - \overline \mu$, $\tilde \theta = \theta - \overline \theta$, and assuming $E = I$, $Q = \overline q I$ with $\overline q \in \mathds{R}$, system $(\ref{alto_complete})$ writes as
\begin{align}\label{alto_complete_lin_example}
\begin{split}
 \left[
       \begin{array}{c}
         \dot{\tilde x}\\
         \dot{\tilde \mu} \\
         \dot{\tilde \theta} \\
       \end{array}
     \right] =& \left[
              \begin{array}{ccc}
                \boldsymbol{0} & -B & I \\
                B^T & \boldsymbol{0} & \boldsymbol{0} \\
                -I & \boldsymbol{0} & -\overline q^2 L^{com} \\
              \end{array}
            \right]\left[
              \begin{array}{c}
                \tilde x \\
                \tilde \mu \\
                \tilde \theta \\
              \end{array}
            \right].
  \end{split}
\end{align}
It can be readily confirmed that
the solution to (\ref{alto_complete_lin_example}), with initial conditions $\tilde{x}({0})=\mathbf{0}$,  $\tilde{\mu}({0})=\mathbf{0}$, and $\tilde{\theta}(0)=\mathds{1}_n $, is given by
\begin{align}
\begin{split}
\tilde{x}(t)&= \mathds{1}_n \sin(t)\\
\tilde{\mu}(t)&=\mathbf{0} \\
\tilde{\theta}(t)&=\mathds{1}_n \cos(t),
\end{split}
\end{align}
which indeed clearly exhibits oscillatory behavior.
\end{remark}

\section{Physical flow dynamics}\label{section:physicalflows}
In the previous discussion we focussed on the design of dynamical flow controllers. On the other hand, flows in networks might follow from underlying physical principles that are not accurately described by (\ref{lambdadotpi}). An important example is the case where the flow $\lambda_k$ directly depends on the states $x_i$ of its adjacent nodes. This is common in {\it e.g.} compartmental systems (see {\it {\it e.g.}} \cite{blanchini_2016_automatica} and \cite{como_2017_arc}).
Another example is when a change of $\lambda$ is induced by the dynamics of the system, instead of a controller that is up to design. We discuss in Subsection
\ref{subpotential} an important example where the flow dynamics are induced by `potential differences'. First we discuss how certain compartmental systems fit within the presented setting.
\subsection{Compartmental systems}\label{subcompt}
Since (\ref{model}) shows similarities with those in compartmental systems, it is natural to wonder how these models are related. Compared to (\ref{model}), compartmental systems  have additional dynamics that model state dependent inflows, outflows and flows between nodes. In this section we incorporate such dynamics in our framework by augmenting (\ref{model}), resulting in
\begin{subequations}\label{model2}
\begin{align}
T_x \dot x &= \Psi(x) -B \lambda + E u - d \label{model_x_alternative}\\
y &=~h(x),
\end{align}
\end{subequations}
where
\be
\Psi(x)=-B_c\gamma(B_c^Th(x))- E_c\eta(E_c^T h(x)).
\ee
Here, $B_c$ is the incidence matrix of a (not necessarily connected) graph $\mathcal{G}_c=(\mathcal{V}, \mathcal{E}_c)$, representing the interconnection of the compartments (\cite{blanchini_2016_automatica}).  Moreover, the set of nodes that have a state dependent inflow/outflow is given by $\mathcal{V}_c \subseteq \mathcal{V}$, with cardinality $p_c := |\mathcal{V}_c| $. Matrix $E_c \in \mathds{R}^{n \times p_c}$ is used to indicate  the locations of the $p_c$ state dependent inflows/outflows and its entries are defined as
\begin{equation*}
\label{eq:incidence}
(e_c)_{ik}=
\begin{cases}
1 \quad &\text{if the $k$-th flow is located at node $i$}\\
0 \quad &\text{otherwise}.
\end{cases}
\end{equation*}
Let $l:=|\mathcal{E}_c|$. The mapping $\gamma(B_c^Th(x)):\R^{l} \rightarrow \R^{l}$ is given by
$\gamma(B_c^Th(x))=(\gamma_1(a_1)\ldots \gamma_{l}(a_{l}))^T$, with $a_k = [B_c^Th(x)]_k$ and $\gamma_k(a_k)$ is nondecreasing and continuously differentiable for all $k \in \mathcal{E}_c$. The mapping $\eta(E_c^T h(x)):\R^{e_c} \rightarrow \R^{e_c}$ is given by $\eta(E_c^T h(x)) = (\eta_1(b_1)\ldots \eta_{e_c}(b_{e_c}))^T$, with $b_i = [E_c^T h(x)]_i$ and $\eta_i(b_i)$ is nondecreasing and continuously differentiable for all $i \in \mathcal{V}_c$.




\begin{remark}[Interpretation of $\Psi(x)$]
In compartmental systems the flow on an edge is often proportional to the potential difference between the two adjacent nodes. Moreover, the inflow or outflow from the system, at a node, is proportional to the potential at the corresponding node (\cite{riaza_2017_jmaa}). The additional term $\Psi(x)$ in (\ref{model2}) models these two situations. More specific, $B_c\gamma(B_c^Th(x))$ models the flow between nodes as a result of potential differences, whereas $E_c\eta(E_c^T h(x))$ models the inflow/outflow. Note that (\ref{model2}) models a compartmental system with additional actuated edges ($e.g.$ flows controlled by a pump) and actuated inputs. The actuation allows us to achieve output regulation and an optimal coordination of the inputs among the nodes, in the presence of unknown disturbances. Previously, in absence of such actuation, works on compartmental systems focussed on proving the asymptotic stability of an arbitrary steady state or required prior knowledge on the disturbance and the network  
(see {\it {\it e.g.} } \cite{blanchini_2016_automatica} and \cite{como_2017_arc}). In some cases, the flow on an edge is proportional to the potential of one of its adjacent nodes (e.g. the flow from a reservoir to another due to gravity). We do not consider this case here and leave the corresponding analysis to a future work.

\end{remark}\stopmodif
%

The optimal control allocation problem \eqref{optimalprod} now becomes
\begin{equation}
\begin{aligned}\label{optimalprod2}
& \underset{u, \lambda}{\text{minimize}}
& & C(u) \\
& \text{subject to}
& & \mathbf{0}=\Psi(\overline x)-B\lambda+E u-d.\\
\end{aligned}
\end{equation}
Similar to Lemma \ref{lemma3}, the following can be immediately shown:
\begin{lemma}[Solution to optimization problem (\ref{optimalprod2})]
The solution to (\ref{optimalprod2}) is given by
\be
\hat{u}  = Q^{-1} (\hat \kappa - r),\label{qpopt2}
\ee
where
\be \label{mu_opt}
  \hat \kappa = E^T \frac{\mathds{1}_n\mathds{1}_n^T }{\mathds{1}_p^T Q^{-1}\mathds{1}_p}(\hat d + E Q^{-1}r ),
\ee
and $\hat d = d + E_c\eta(E_c^T h(\overline x))$.

\end{lemma}
%
Due to the new network dynamics (\ref{model2}) and optimal control input $\hat u$ in the network, Assumption \ref{assump_bounded_controller_outputs2} needs to be revisited.
\begin{assumption}[Attainability revisited] \label{assump:feasibility_rev}
 ~Consider functions $f_k(\cdot)$ and $g_i(\cdot)$, in respectively (\ref{lambdadotpi}) and (\ref{controllerinput}).
 Let $\hat{u}$ be as in (\ref{qpopt2}).  There exists a $\omega\in\mathds{R}^m$, such that $[B^\dagger (\Psi(\overline x) + E\hat u -d) +(I-B^\dagger B) \omega]_k \in \mathcal{R}(f_k)$  for all $k\in\mathcal{E}$. Furthermore, $\hat u_i\in \mathcal{R}(g_i)$ for all $i\in\mathcal{V}_e$.
\end{assumption}
With the assumption above, we can prove, similarly as Lemma \ref{lem:steady_state},  the existence of a steady state for system (\ref{lambdadotpi}), (\ref{controllerinput}), (\ref{model2}).
 The argumentation is along the lines of the proof of Lemma \ref{lem:steady_state} and we omit the details. We can now prove the following result:
\stopmodif
\begin{theorem}[Solving Problem 1 for system (\ref{model2})]\label{theorem2}
 Let Assumptions  \ref{connected_graph_dynamics}-- \ref{assump:com} and \ref{assump_bounded_controller_outputs}--\ref{assump:feasibility_rev} hold.  The solutions to system (\ref{model2}), in closed loop with (\ref{lambdadotpi}) and (\ref{controllerinput}), globally converge to point in the set
 \be
 \label{seUpsilon_2}
\Upsilon_2=\left\{ x, \mu, \xi, \theta, \phi \left| \begin{tabular}{l}
                                                      $B(f(\mu)-f(\overline \mu))=\boldsymbol{0}$,  \\
                                                       $B(\xi-\overline \xi)=\boldsymbol{0}$,  \\
                                                      $x=\overline x$, $\theta=\overline \theta$, $\phi=\overline \phi$\\
                                                     \end{tabular} \right.\right\},
\ee
where  $\lambda = f(\mu)$  is a constant, $h(x) = \overline y$ and where  $u = g(\theta)  = \hat u$, with $\hat u$ given by (\ref{qpopt2}).
Moreover, $u=g(\theta)$ and $\lambda=f(\mu)$ satisfy constraints (\ref{upsatbound}) and (\ref{uesatbound}) $\text{for all } t \geq 0$. Therefore, controllers (\ref{lambdadotpi}) and (\ref{controllerinput}) solve Problem \ref{problem1} for the flow network (\ref{model2}).

\end{theorem}
\textbf{Proof.}
First, the fulfilment of the the constraints (\ref{upsatbound}) and (\ref{uesatbound}) $\text{for all } t \geq 0$ is guaranteed by the design of the controllers. Second, a straightforward adjustment of the arguments of Theorem \ref{theorem1} shows that the same incremental storage function (\ref{storagefunction}), used in Theorem \ref{theorem1}, now satisfies
\be \label{vdot_alternative}
\begin{aligned}
\dot{V}(\cdot)  =& ~(h( x)-h(\overline x))^T(\Psi(x)-\Psi(\overline x)) \\
& ~-  (\phi -\overline \phi)^T  QL^{com}Q (\phi-\overline \phi) \\
   & ~ - (g(\theta) - \phi)^T (g(\theta) - \phi) \\
   &~ -(f(\mu)-\xi)^T(f(\mu)-\xi),
\end{aligned}
\ee
along the solutions to (\ref{model2}) in closed loop with (\ref{lambdadotpi}) and (\ref{controllerinput}). We continue by showing that the additional term in $\dot V(\cdot)$  (comparing with the expression of $\dot{V}(\cdot)$ in (\ref{lemma_v_dot})) satisfies
\be \label{psy_leq_zero}
(h( x)-h(\overline x))^T(\Psi(x)-\Psi(\overline x))\leq 0.
\ee
In fact, since $\gamma(\cdot)$ and $\eta(\cdot)$ are increasing mappings and by application of Hadamard's lemma we have that
\begin{align}\label{vdotthereom2}
  & ~(h( x)-h(\overline x))^T(\Psi(x)-\Psi(\overline x)) \\
  =& ~-(h( x)-h(\overline x))^TB_c(\gamma(B_c^Th(x))-\gamma(B_c^Th(\overline x)) \nonumber \\
  &~-(h( x)-h(\overline x))^T E_c(\eta(E_c^Th(x)) -\eta(E_c^Th(\overline x)))  \nonumber \\
  =&~-(h( x)-h(\overline x))^TB_c \Gamma^b(x) B_c^T(h(x)- h(\overline x))  \nonumber \\
  &~-(h( x)-h(\overline x))^T E_c \Gamma^e(x) E_c^T(h(x)- h(\overline x)) \leq 0, \nonumber
\end{align}
where $\Gamma^b(x)$ and $\Gamma^e(x)$ are diagonal matrices with entries
\begin{align}
{\Gamma}^b_{kk}(x)&=\int_0^1 \left.\frac{\partial \gamma_k(y_k)}{\partial y_k} \right|_{y_k=\tau (\chi^b_k(x)-\chi^b_k(\overline x))+\chi^b_k(\overline x)} d\tau\\
{\Gamma}^e_{ii}(x)&=\int_0^1 \left.\frac{\partial \eta_i(y_i)}{\partial y_i} \right|_{y_i=\tau (\chi^e_i(x)-\chi^e_i(\overline x))+\chi^e_i(\overline x)} d\tau,
\end{align}
with $\chi^b_k(x)=[B_c^Th( x)]_k$ and $\chi^e_i( x)=[E_c^Th( x)]_i$.
Note that ${\Gamma}^b_{kk}(x),{\Gamma}^e_{ii}(x) \geq 0$ for any $x$, since $\gamma_k(\cdot)$ and $\eta_i(\cdot)$ are increasing functions for all $k \in \mathcal{E}_c$ and all $i \in \mathcal{V}_c$.
 Therefore, $V(\cdot)$ satisfies
 \be \label{vdot_alternative2}
\begin{aligned}
\dot{V}(\cdot)  \leq & ~-  (\phi -\overline \phi)^T  QL^{com}Q (\phi-\overline \phi) \\
   & ~ - (g(\theta) - \phi)^T (g(\theta) - \phi) \\
   &~ -(f(\mu)-\xi)^T(f(\mu)-\xi),
\end{aligned}
\ee
along the solutions to (\ref{model2}) in closed loop with (\ref{lambdadotpi}) and (\ref{controllerinput}). Note that expression (\ref{vdot_alternative2}) is identical to (\ref{lemma_v_dot}), that is used to prove Theorem 1 above. Similar to the proof of Theorem \ref{theorem1}, we can argue that $x = \overline x$, by exploiting the relations (\ref{invariantset1mu}) --  (\ref{invariantset1phi}). Therefore, on the invariant set where $\dot V(\cdot) = 0$,
\be
T_x \dot x =\Psi(x)-\Psi(\overline x)-B (f(\mu)-f(\overline \mu)) + E( \phi- \overline \phi),
\ee
reduces to (\ref{invariantset1x}), such that system (\ref{model2}) in closed loop with (\ref{lambdadotpi}) and (\ref{controllerinput}), is on the invariant set identical to (\ref{invariantset1}).
 From here, the proof follows the same steps as the proof of Theorem \ref{theorem1}.
\qedp

\subsection{Potential induced flow dynamics}\label{subpotential}
%
%
%
In this subsection we study a network where the flow dynamics are given by the following expression:
\begin{align} \label{lambdadotpi_alternative}
\begin{split}
T_\mu \dot \mu =& ~   B^T (h(x) - \overline y)  \\
\lambda=& ~f(\mu),
  \end{split}
\end{align}
that has been studied in the context of networked systems in {\it {\it e.g.} } \cite{van2012hamiltonian}, \cite{burger2014duality} and \cite{burger_2015_tcns}. Also, it describes the behaviour of inductive lines in an electric network (see the case study on a multi-terminal HVDC network in Subsection \ref{sec:hvdc}).

The dynamics (\ref{lambdadotpi_alternative}) coincide with (\ref{lambdadotpi}),  if one neglects the terms depending on the now missing state $\xi$.
In fact, (\ref{lambdadotpi_alternative}) can generate the same steady state output as (\ref{lambdadotpi}) and also shares an incremental passivity property. However, as we pointed out in Remark \ref{rem:avoiding_oscillation}, the state $\xi$ is essential to derive the convergence result in Theorem 1. On the other hand, by carefully selecting  nodes that have a controllable external input,  the controllers (\ref{controllerinput}) and (\ref{lambdadotpi_alternative}) still solve Problem 1 for the flow network (\ref{model}).
This choice is based on the notion of a zero forcing set (see {\it e.g.}, \cite{hogben_2010_laa, monshizadeh2014zero, trefois_2015_laa}), which we review next.

Consider the graph  $\mathcal{G}$ and let us initially color each of its nodes either black or white. The color of the nodes then changes according to the following coloring rule:

\medskip

\noindent {\bf Graph coloring rule} {\it If node $i$ is colored black and has exactly one neighbor $j$ which is white, then  the color of node $j$ is changed to black.}

\medskip

Let $\mathcal{V}_0 \subseteq \mathcal{V}$ be the set of nodes which are initially colored black, while the remaining ones are white, and let $\mathcal{C}(\mathcal{V}_0)$ be the set of black node obtained by applying the color changing rule until no more changes are possible.
A zero forcing set is then defined as:
\begin{definition}[Zero forcing set]
If $\mathcal{V}_0\subseteq\mathcal{V}$ satisfies $\mathcal{C}(\mathcal{V}_0)=\mathcal{V}$ then $\mathcal{V}_0$ is a zero forcing set for $\mathcal{G}$.
\end{definition}

We now make a connection between a zero forcing set and the set  $\mathcal{V}_{e}$ of nodes that have actuation ({\it {\it i.e.},} all nodes that correspond to the rows of $E$ that contain a non-zero entry).
\begin{assumption}[$\mathcal{V}_e$ is a zero forcing set]\label{as:zero_forcing set}
The set $\mathcal{V}_e$ is a zero forcing set for $\mathcal{G}$.
\end{assumption}
An example of a zero forcing set is provided previously Figure \ref{communication_graph}, where the black nodes indeed form a zero forcing set for the physical network.

We are now ready to state the second result of this section.
\begin{theorem}[Solving Problem 1 with (\ref{lambdadotpi_alternative})]\label{theorem3}
%
Let Assumptions \ref{connected_graph_dynamics}--\ref{assump_bounded_controller_outputs} and \ref{as:zero_forcing set} hold. The solutions to system (\ref{model}),  in closed loop with the controllers (\ref{controllerinput}) and (\ref{lambdadotpi_alternative}), globally converge to a point in the set
\be
 \label{setUpsilon_3}
\Upsilon_3=\left\{ x, \mu, \theta, \phi \left| \begin{tabular}{l}
                                                      $B(f(\mu)-f(\overline \mu))=\boldsymbol{0}$,  \\
                                                      $x=\overline x$, $\theta=\overline \theta$, $\phi=\overline \phi$\\
                                                     \end{tabular} \right.\right\},
\ee
where $\lambda = f(\mu)$  is a constant, $h(x) = \overline y$ and where  $u = g(\theta)  = \overline u$, with $\overline u$ given by (\ref{qpopt}).
Moreover, $u=g(\theta)$ and $\lambda=f(\mu)$ satisfy constraints (\ref{upsatbound}) and (\ref{uesatbound}) $\text{for all } t \geq 0$. Therefore, controllers (\ref{controllerinput}) and (\ref{lambdadotpi_alternative}) solve Problem \ref{problem1} for the flow network (\ref{model}).
\end{theorem}

\textbf{Proof.}
Following the argumentation of the proof of Theorem \ref{theorem1}, using the same incremental storage function (\ref{storagefunction}), allows us to conclude  that the
solutions to the  system  (\ref{model}), (\ref{controllerinput}),  (\ref{lambdadotpi_alternative}) approach the largest invariant set contained in the set where $\dot V(\cdot)=0$. This set, where  $\dot V(\cdot)=0$,
is now characterized by
\be \label{setS_alternative}
\mathcal{S}_3=\left\{ x, \mu, \theta, \phi \left|
                                                      \phi = g(\theta), Q(\phi -\overline \phi)  \in \text{im}(\mathds{1}_p)\right.\right\}.
\ee
System  (\ref{model}), (\ref{controllerinput}),  (\ref{lambdadotpi_alternative}) satisfies on this set
\begin{subequations}\label{invariantset1_alternative}
 \begin{align}
T_x \dot x =&-B (f(\mu)-f(\overline \mu))+E(\phi-\overline \phi)  \label{invariantset1_alternative1}\\
T_\mu  \dot \mu =& ~    B^T (h(x) - h(\overline x))  \label{invariantset1_alternative2}\\
\mathbf{0} =&  -E^T(h(x) - h(\overline x)) \label{invariantset1_alternative3}\\
T_\phi  \dot \phi=&~\mathbf{0}. \label{invariantset1_alternative4}
\end{align}
\end{subequations}
We now prove by induction that $h_i(x_i) = h_i(\overline x_i)$ for all $i\in \mathcal{V}$. To this end, let us define the  sequence of sets of nodes $\mathcal{V}_k\subseteq \mathcal{V}$, with $k\in\mathds{N}_{\geq 0}$, having the properties:
\smallskip \\
(i) $\mathcal{V}_k$ is a zero forcing set; \smallskip \\
(ii) on the largest invariant set for (\ref{model}), (\ref{controllerinput}), (\ref{lambdadotpi_alternative}) contained in $\mathcal{S}_3$, it holds that $h_i(x_i)=h_i(\overline x_i)$ for all $i\in \mathcal{V}_k$.
\smallskip \\
Let the cardinality of $\mathcal{V}_k$ be denoted by $n_k$. In order to show that $h_i(x_i) = h_i(\overline x_i)$ for all $i\in\mathcal{V}$ we will prove that there exists an index $\overline k$ such that $n_{\overline k}=n$, where $\mathcal{V}_{\overline k}$ satisfies properties $(i)$ and $(ii)$. Recall that $|\mathcal{V}| = n$.

First, we note that Assumption \ref{as:zero_forcing set} and (\ref{invariantset1_alternative3}) imply that $\mathcal{V}_e$ satisfies properties (i) and (ii). For this reason, we can set $\mathcal{V}_0:=\mathcal{V}_e$ and $n_0:=p>0$ that satisfies properties (i) and (ii). If $n_0=n$, then $\overline k=0$, otherwise $n_0<n$ and we proceed as follows.

For a $k\in\mathds{N}_{\geq0}$, we consider a set of nodes  $\mathcal{V}_k$ of cardinality $0<n_k<n$ satisfying properties (i) and (ii) above. We will show that this implies that there exists a set of nodes  $\mathcal{V}_{k+1}$ that satisfies properties (i) and (ii) with $n_k<n_{k+1}$.

Let us define
\[
B^{(k)}=\left[
    \begin{array}{c}
      B^{\mathcal{B}(k)} \\
      B^{\mathcal{W}(k)} \\
    \end{array}
  \right],
\]
where the matrices $B^{\mathcal{B}(k)} \in \mathds{R}^{n_k \times m}$ and $B^{\mathcal{W}(k)} \in \mathds{R}^{(n-n_k) \times m}$ are obtained by collecting from $B$ the rows indexed by $\mathcal{V}_k$ and $\mathcal{V}\backslash \mathcal{V}_k$, respectively. Note that $B^{(k)}$ is obtained from $B$ by reordering of the rows, and that $B^{\mathcal{B}(k)}$ and $B^{\mathcal{W}(k)}$  are the rows of $B$ corresponding to the black and white nodes, respectively.
Similarly, for any vector $\chi\in\mathds{R}^n$ let $\chi^{\mathcal{B}(k)}\in \mathds{R}^{n_k}$ and $\chi^{\mathcal{W}(k)}\in \mathds{R}^{n-n_k}$ be obtained by collecting from $\chi$ the elements indexed by $\mathcal{V}_k$ and $\mathcal{V}\backslash \mathcal{V}_k$ respectively.
%
%
We note that, by   property (ii), on the largest invariant set, the set $\mathcal{V}_k$ fulfils $(h(x)-h(\overline x))^{\mathcal{B}(k)}=\mathbf{0}$.
More explicitly, $h_i(x_i)-h_i(\overline x_i)=0$ for all $i\in \mathcal{V}_k$. By the strict monotonicity of $h_i(x_i)$, it follows that on the invariant set $x_i=\overline x_i$ for all $i \in \mathcal{V}_k$. Since $\frac{d}{dt}( \phi-\overline \phi)=\boldsymbol{0}$ due to (\ref{invariantset1_alternative4}), on the invariant set we have,  by \eqref{invariantset1_alternative1} and  (\ref{invariantset1_alternative2}), that
\begin{equation*}
\begin{aligned}
&T_x \left[
  \begin{array}{c}
    \mathbf{0} \\
    \ddot{x}^{{\mathcal{W}(k)}} \\
  \end{array}
\right]=-B^{(k)}\frac{\partial f(\mu)}{\partial \mu}{B^{(k)}}^T \left[
                                          \begin{array}{c}
                                            \mathbf{0}  \\
                                            (h(x)-h(\overline x))^{{\mathcal{W}(k)}}
                                          \end{array}
                                        \right],
\end{aligned}
\end{equation*}
from which it follows that
\be\label{sethxiszero2}
\mathbf{0}=-B^{{\mathcal{B}(k)}} \frac{\partial f(\mu)}{\partial \mu} {B^{{\mathcal{W}(k)}}}^T (h(x)-h(\overline x))^{{\mathcal{W}(k)}}.
\ee
Note that $B^{{\mathcal{B}(k)}} \frac{\partial f(\mu)}{\partial \mu} {B^{{\mathcal{W}(k)}}}^T$ is the right-upper block of the Laplacian matrix $B^{(k)}\frac{\partial f(\mu)}{\partial \mu}{B^{(k)}}^T$ with strictly positive weight matrix, since $f_k(\mu_k)$ is strictly increasing, such that $\frac{\partial f_k(\mu_k)}{\partial \mu_k}>0$ for all $k\in\mathcal{E}$. The non-zero entries in $B^{{\mathcal{B}(k)}} \frac{\partial f(\mu)}{\partial \mu} {B^{{\mathcal{W}(k)}}}^T$
correspond to pairs of exactly one black and one white node that are connected via an edge. Therefore we have that each row $i$ of $B^{{\mathcal{B}(k)}} \frac{\partial f(\mu)}{\partial \mu} {B^{{\mathcal{W}(k)}}}^T$ (which corresponds to a black node) contains a strictly negative number at entry $j$ if, and only if, node $n_k+j$ is a neighbor of the node $i$. By assumption we have that $\mathcal{V}_k$ is a zero forcing set and that $\mathcal{V}_k\subsetneq \mathcal{V}$, which implies that there exists at least one row of $B^{{\mathcal{B}(k)}} \frac{\partial f(\mu)}{\partial \mu} {B^{{\mathcal{W}(k)}}}^T$ which contains exactly one non-zero entry.
Let $\mathcal{U}_k$ be the set in which we collect the nodes that correspond to these rows and define $\mathcal{V}_{k+1}:=\mathcal{V}_k\cup \mathcal{U}_k$.
From (\ref{sethxiszero2}), we  have that
$
0=h_i(x_i)-h_{i}(\overline x_i),
$
for all $i\in\mathcal{U}_k$ and therefore for all $i\in\mathcal{V}_{k+1}$. Moreover, since $\mathcal{V}_{k}\subset\mathcal{V}_{k+1}$, and since we assume that $\mathcal{V}_{k}$ is a zero forcing set for $\mathcal{G}$, also $\mathcal{V}_{k+1}$ is a zero forcing set for $\mathcal{G}$. This concludes the proof that there exists $\mathcal{V}_{k+1}$ that satisfies properties (i) and (ii), with $n_{k+1} > n_k$.

Since the number of nodes is finite,  in a finite number of iterations $\overline k$ we arrive at a set $\mathcal{V}_{\overline k}$ where $n_{\overline k}=n$, {\it {\it i.e.} } $\mathcal{V}_{\overline k}$ coincides with $\mathcal{V}$ and  has the property that
on the largest invariant set for
(\ref{model}), (\ref{controllerinput}), (\ref{lambdadotpi_alternative}) contained in $\mathcal{S}_3$,
$0=h_i(x_i)-h_{i}(\overline x_i)$ for all $i\in \mathcal{V}$.  From here, omitting the variable $\xi$, the proof follows, \emph{mutatis mutandis}, the proof of Theorem \ref{theorem1}, starting from the paragraph below (\ref{parastart})
 \\  \stopmodif
\qedp
\begin{remark}[Relaxing Assumption \ref{as:zero_forcing set}]
In the case that $f(\mu)=\mu$ and $h(x)=x$, successive differentiations of  (\ref{invariantset1_alternative3}) yields
\begin{align}
  \boldsymbol{0} = \underbrace{\begin{bmatrix} -E^T  \\
   E^T Y  \\
   -E^T Y^2\\
   \vdots \\
   (-1)^{n}E^TY^{n-1}
   \end{bmatrix}}_{\mathcal{O}}(h(x)- h(\overline x)),
\end{align}
where $Y = T^{-1}_{x}BT^{-1}_{\mu}B^T$.
To conclude that $h(x) = h(\overline x)$, it is sufficient that the matrix $\mathcal{O}$ has full column rank, {\it i.e.} the pair $(E^T, Y)$ is observable. Although, a similar argumentation can be performed with the nonlinear mappings $f(\mu)$ and $h(x)$, it does not immediately lead to a simple criterion that permits to conclude $h(x) = h(\overline x)$.
\end{remark}
After separately discussing the particular modifications to the flow network and controllers in Subsections \ref{subcompt} and \ref{subpotential}, we briefly discuss the combination of both in the corollary below:
\begin{corollary}[Combined modifications]
 Let Assumptions  \ref{connected_graph_dynamics}-- \ref{assump:com} and \ref{assump_bounded_controller_outputs}--\ref{as:zero_forcing set} hold. Consider the flow network (\ref{model2}) and let $\mathcal{V}_s \subseteq \mathcal{V}_c$, be defined as\footnote{In Theorem \ref{theorem2}, we only required $\eta_i(y_i)$ to be nondecreasing for all $i \in \mathcal{V}_c$, \emph{i.e.} $\frac{\partial \eta_i(y_i)}{\partial y_i} \geq 0$.}
 \begin{align}
 \begin{split}
 \mathcal{V}_s =  \{i \in \mathcal{V}_c |~ \frac{\eta_i(y_i)}{\partial y_i}\Big|_{y_i=[E_c^T h(\overline x)]_i} >0   \}.
 \end{split}
 \end{align}
 If $\mathcal{V}_e \cup \mathcal{V}_s$ is a zero forcing set for $\mathcal{G}$, then the solutions to system (\ref{model2}),  in closed loop with the controllers (\ref{controllerinput}) and (\ref{lambdadotpi_alternative}), globally converge to a point in the set
\be
 \label{setUpsilon_4}
\Upsilon_4=\left\{ x, \mu, \theta, \phi \left| \begin{tabular}{l}
                                                      $B(f(\mu)-f(\overline \mu))=\boldsymbol{0}$,  \\
                                                      $x=\overline x$, $\theta=\overline \theta$, $\phi=\overline \phi$\\
                                                     \end{tabular} \right.\right\},
\ee
where $\lambda = f(\mu)$  is a constant, $h(x) = \overline y$ and where  $u = g(\theta)  = \overline u$, with $\overline u$ given by (\ref{qpopt}). Therefore, controllers (\ref{controllerinput}) and (\ref{lambdadotpi_alternative}) solve Problem \ref{problem1} for the flow network (\ref{model2}).
\end{corollary}

\textbf{Proof.}
  Following a similar argumentation as in the the proof of Theorem \ref{theorem3}, $x_i = \overline x_i$ for all $i \in \mathcal{V}_e$. Moreover, the dynamics (\ref{model2}) give rise to an additional term in $\dot V(\cdot)$ in the same manner as in the proof of Theorem \ref{theorem2} (see (\ref{vdotthereom2})), namely:
  \begin{align}
  -(h( x)-h(\overline x))^T E_c \Gamma^e(x) E_c^T(h(x)- h(\overline x)) < 0.
  \end{align}
  Consequently, on the largest invariant set where $\dot V(\cdot)=0$, also
   $x_i = \overline x_i$ for all $i \in \mathcal{V}_s$, since $\eta_i(y_i)$ is strictly increasing around $[E_c^T h(\overline x)]_i$, for all $i \in \mathcal{V}_s$. From here the proof continues along the lines of the proof of Theorem \ref{theorem3}.
\qedp
\section{Case studies} \label{case_study}
To illustrate how physical systems can be regarded as a flow network and to show the performance of the proposed controllers we consider two case studies. The first case study considers a district heating system, whereas the second case study considers a multi-terminal high voltage direct current (HVDC) network.
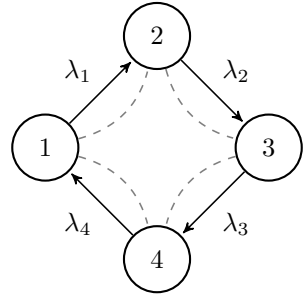
\begin{figure}
\begin{center}
\begin{tikzpicture}[>=stealth',shorten >=1pt,auto,node distance=2.1cm,
                    semithick]
  \tikzstyle{every state}=[circle,thick,fill=white!10,draw=black,text=black]

 \node[state] (A)                    {1};
  \node[state]         (B) [above right of=A] {2};
  \node[state]         (D) [below right of=A] {4};
  \node[state]         (C) [below right of=B] {3};

  \path[->] (A) edge             node {$\lambda_{1}$} (B)
  		     (D)  edge		     node {$\lambda_{4}$} (A)
           (B) edge              node {$\lambda_{2}$} (C)
           (C) edge              node {$\lambda_{3}$} (D);

  \path[] (A) edge [bend right, dashed, gray]          node {} (B)
  		edge	 [bend left, dashed, gray]	     	node {} (D)
           (B) edge [bend right, dashed, gray]          	node {} (C)
           (C) edge [bend right, dashed, gray]          	node {} (D);
\end{tikzpicture}
\caption{Topology of the considered heat network. The arrows indicate the required flow directions in the heat network, while the dashed lines represent the communication network used by the controllers.}
\label{example_graph}
\end{center}
\end{figure}

\subsection{District heating system}
Continuing our previous work in \cite{Scholten2015}, we consider a district heating system with a topology as depicted in Fig \ref{example_graph}.
%
%
%
\begin{figure}
\centering
  \includegraphics[trim=1cm 12cm 17cm 0.5cm, clip=true, width=0.37 \textwidth]{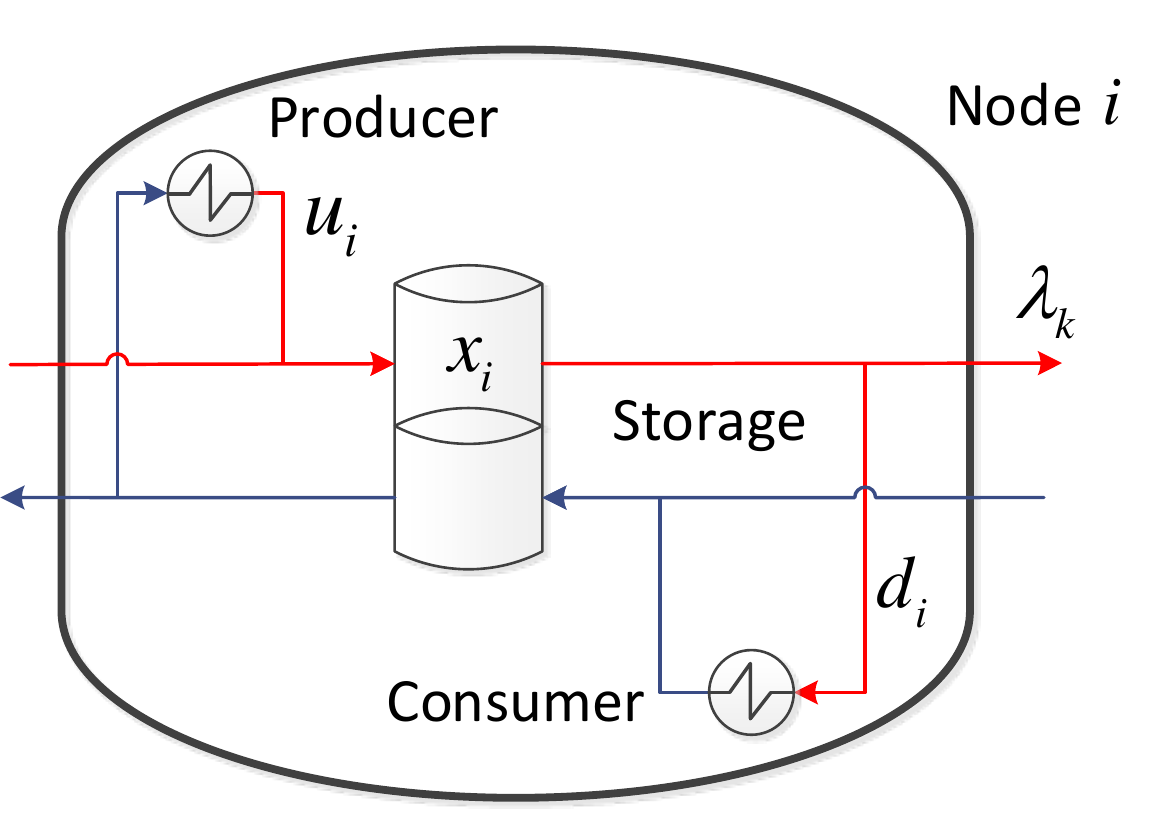}\\
  \caption{A node in the district heating network.}\label{node}
\end{figure}
 Each node represents a producer, a consumer and a stratified storage tank (see Fig. \ref{node}). The storage tank consists of a hot and a cold layer of water, both with variable volumes. We denote the volume of the hot layer of water at node $i$ as $x_i$ ($m^3$), which is also the measured output of the system, {\it i.e.} $h_i(x_i) = x_i$.  The various nodes are interconnected via a pipe network $\mathcal{G}$. Following \cite{Scholten2015}, the dynamics for the hot layer can be derived by applying mass conservation laws resulting in the following representation of the district heating system:
\begin{align}
\dot{x}&=-B\lambda+u-d, \label{dynamicsstoragehot1}
\end{align}
where $\lambda_k$ ($m^3/h$) denotes the flow through pipe $k$. Moreover, $u_i$ ($m^3/h$) and $d_i$ ($m^3/h$) are respectively the flow trough the heat exchanger of the producer and the consumer at node $i$. It is immediate to see that (\ref{dynamicsstoragehot1}) has identical dynamics as (\ref{model}) if we set $T_x=I$. The controllers (\ref{lambdadotpi}) and (\ref{controllerinput}) are therefore applicable and we study the obtained closed-loop system.
\begin{figure}
\centering
\includegraphics[trim=3cm 6.5cm 3cm 6.5cm, clip=true, width= 0.48 \textwidth]{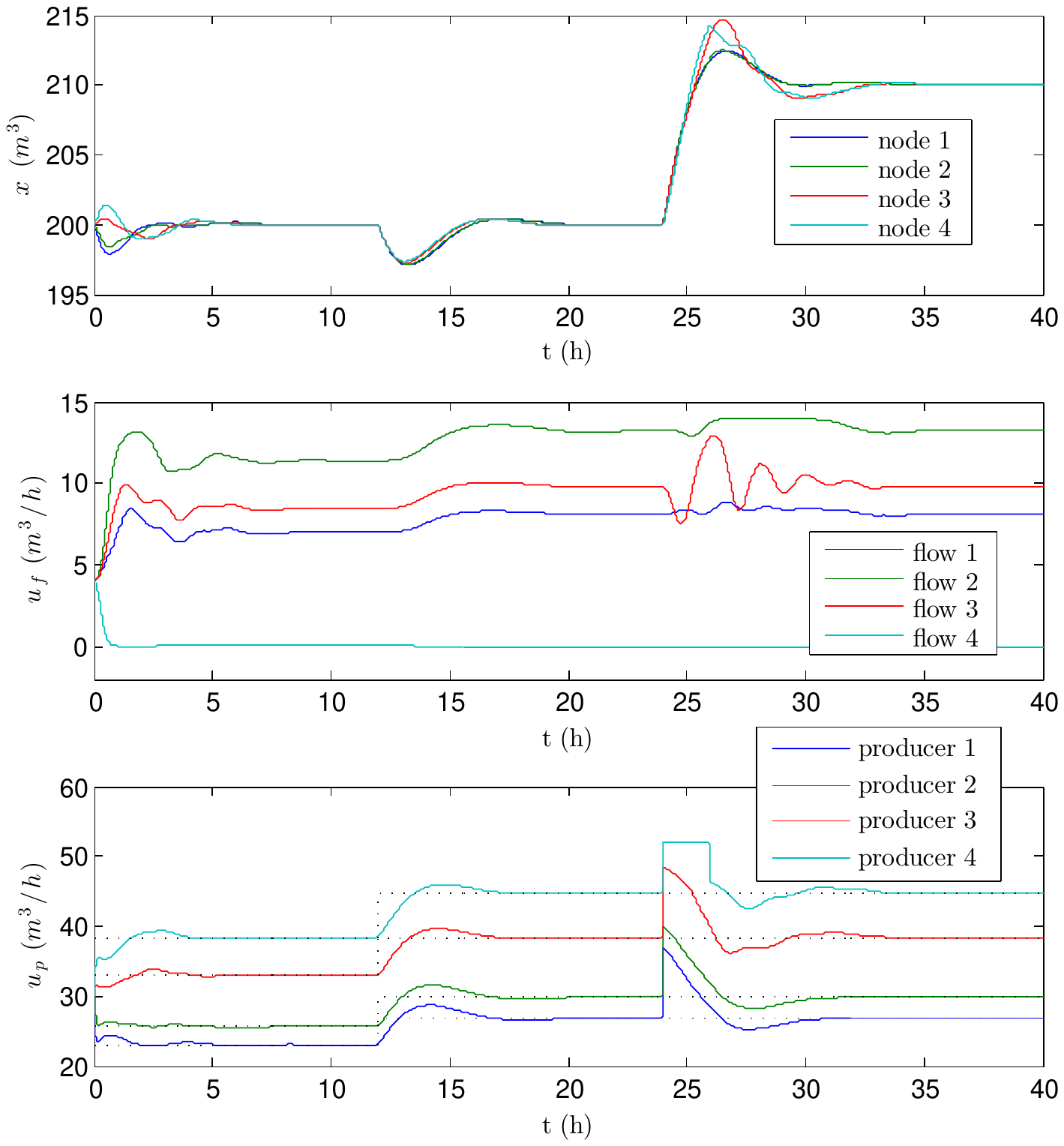}\\
\caption{Volumes, flows and productions of the district heating system during a 40 hour period. The optimal production $\overline u_p$ as in (\ref{qpopt}) is indicated by dotted lines in the lower plot. }\label{case_study1}
\end{figure}

We perform a simulation over a $40$ hours time interval in which we evaluate the response to a change in demand at $t=12$ and change in setpoint at $t=24$. The cost functions of the four producers are purely quadratic, {\it i.e.} $s = r =\mathbf{0}$. We take
$Q=\text{diag}\left(
  \begin{array}{cccc}
   10 & 9& 7& 6 \\
  \end{array}
\right)$.
Initially the volume is
$x(0) =\begin{bmatrix}
   200 & 200 & 200 & 200 \end{bmatrix}^T$,  which is also to the  setpoint $\overline x (t)$ for all $t< 24$. The initial demand is given by
$d(t)=\begin{bmatrix}
   30 &30& 30& 30 \\
  \end{bmatrix}^T$, for all $t<12$, which is increased to $d(t)=\begin{bmatrix}
   35 &35& 35& 35 \\
   \end{bmatrix}^T$,
for all $t \geq 12$. The setpoint for the volume $\overline x(t)$ is increased at $t=24$ to
$\overline x(t)=\begin{bmatrix}
   210 & 210 & 210 & 210
  \end{bmatrix}^T$, for all $t \geq 24$. To guarantee uni-directional flows and positive production we require $\lambda_k>0$ and $u_i>0$, for all $k,i \in \{1,2,3,4 \}$. Due to capacity constraints, we additional require them to be upper bounded by $14$ $m^3/h$ and $52$ $m^3/h$, respectively. To enforce these constraints, the output of the controllers is designed as
\begin{align}
\begin{split}
\lambda_{k}&= f_k(\mu_k)=7(\tanh(\mu_k)+1) \\
u_{i} &= g_i(\theta_i)=26(\tanh(\theta_i)+1),
\end{split}
\end{align}
where $\tanh(\cdot)$ is the hyperbolic tangent function. 
Finally, we let $T_\mu=I$, $T_\theta=I$, $T_\phi=0.005\cdot I$ and we set all the weights of $L^{com}$ to $10$ and we let it be undirected which implies that $L^{com}$ is balanced.

The resulting response of the system can be found in Figure \ref{case_study1}, where we can clearly see the effects of the increased demand at $t=12$ and change in setpoint at $t=24$. More specifically, in the upper plot we can see that the controllers indeed let the volumes in the four storage tanks to converge towards the desired setpoints of $200m^3$ ($t < 24$) and $210m^3$ ($t \geq 24$). In the middle plot we see that the flows in the pipes remain within the constraint $0<\lambda_k<14$ for all $k \in \{1,2,3,4 \}$ throughout the entire simulation. Finally, in the bottom plot, the production at the four nodes is given, where the optimal productions is denoted by the dotted lines. We observe that the production converges towards the optimal value $\overline u$ and satisfies $0<u_i<52$ for all $i \in \{1,2,3,4 \}$, during the entire simulation period.
\stopmodif
\subsection{Multi-terminal HVDC networks}\label{sec:hvdc}
\begin{figure}
\centering
\begin{circuitikz}[american currents,scale=0.8, transform shape] \draw
 (1,2) node[anchor = north west]{$V_1$}
  (3,2) node[anchor=north west]{$V_2$}
   (1,0) node[anchor=south west]{$V_4$}
    (3,0) node[anchor=south west]{$V_3$}
  (1,0)to[L, l= ] (1,2)
   (1,2)to [L, l_=] (3,2)
  (3,0) to[L, l=] (3,2)
 (3,0) to[L, l_= ]  (1,0)
 (-2,2) to[I, i_>=$ d_1 $] (1,2)
  (-2,0) to[I, i_>=$u_{4} - d_4 $] (1,0)
   (6,2) to[I, i^>=$u_{2} - d_2 $] (3,2)
  (6,0) to[I, i^>=$u_{3} - d_3 $] (3,0)
(-2,0) -- (-2,-1) node[ground]{}
(6,0) -- (6,-1) node[ground]{}
(-2,2) -- (-2,3) node[ground, rotate=180]{}
(1,3) -- (1,3) node[ground, rotate=180]{}
(3,3) -- (3,3) node[ground, rotate=180]{}
(6,2) -- (6,3) node[ground, rotate=180]{}
(1,2) to[capacitor={}](1,3)
(3,2) to[capacitor={}](3,3)
(1,0) to[capacitor={}](1,-1)
(3,0) to[capacitor={}](3,-1)
(3,-1) -- (3,-1) node[ground]{}
(1,-1) -- (1,-1) node[ground]{}
;\end{circuitikz}
\caption{Topology of a four bus multi-terminal HVDC network. We take $\mathcal{C}_i = 57 \mu F$ and $\mathcal{L}_k= 0.0135 H$ for $i,k \in \{1, 2,3, 4\}$. } \label{hvdc network}
\end{figure}
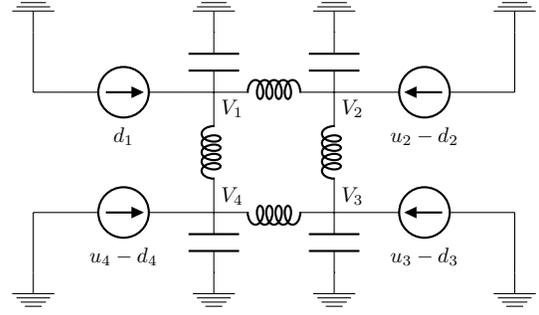

\begin{figure}
\centering
\includegraphics[trim=3.8cm 8.2cm 3.8cm 8.2cm, clip=true, width= \columnwidth]{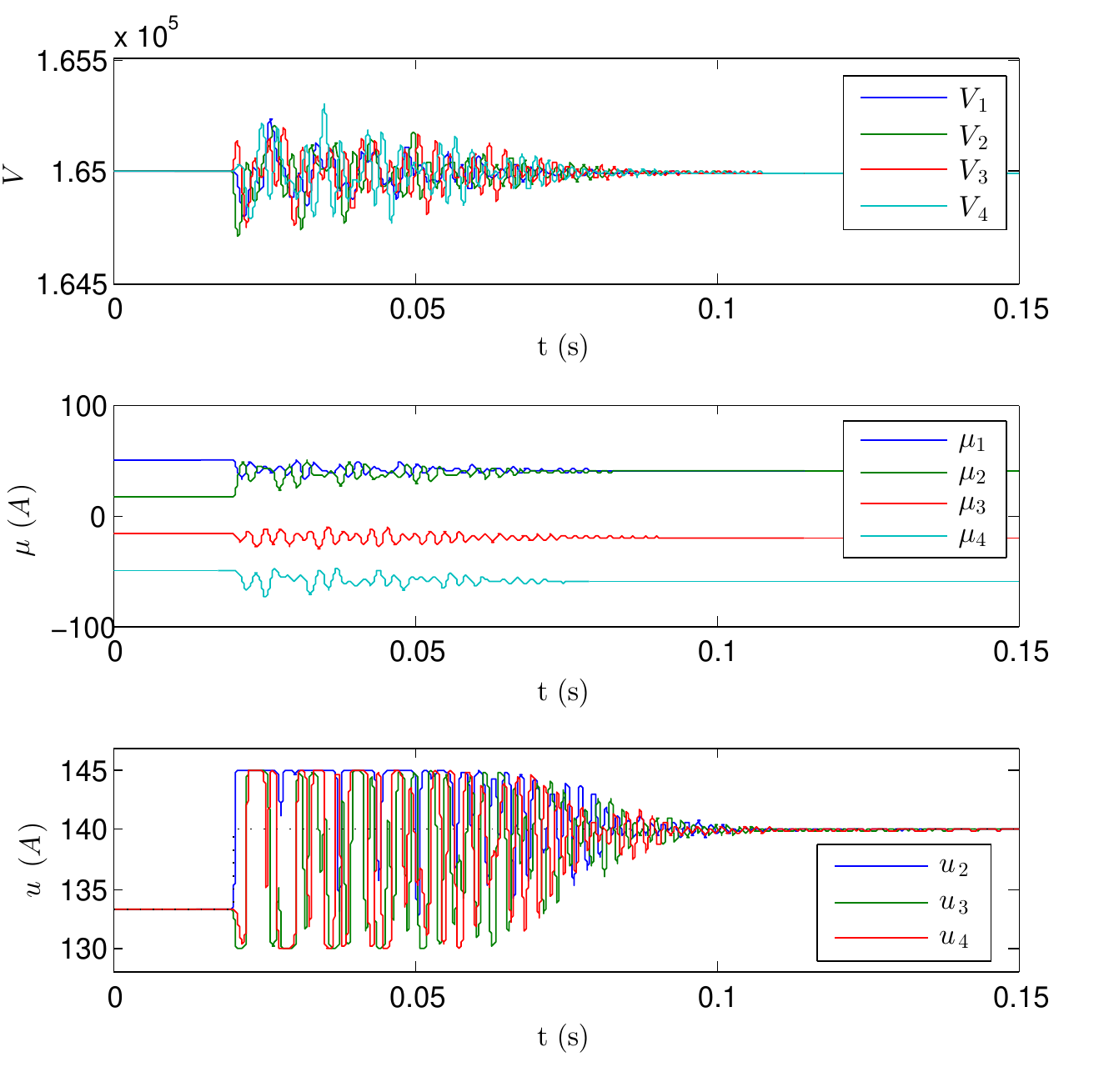}\\
\caption{Voltages, current flows and current injections for a high voltage direct current network. The optimal production $\overline u_p$ as in (\ref{qpopt}) is indicated by dotted lines in the lower plot. }\label{case_study2}
\end{figure}

As a second case study we consider multi-terminal high voltage direct current (HVDC) networks that have been recently studied in {\it e.g.} \cite{zonetti_2015_cep} and \cite{andreasson_2016_tns}. We assume that the lines connecting the terminals are lossless, such that the overall network dynamics are given by
\begin{align}
\begin{split}\label{hvdcmodel}
  \mathcal{C} \dot V =& -B\mu +u - d \\
  \mathcal{L} \dot \mu =&~ B^TV,
  \end{split}
\end{align}
where $V$ are the voltages at the terminals, $\mu$ are the currents through the lines, $d$ are \emph{uncontrollable} current loads and $u$ are the \emph{controllable} current injections. We consider a circuit of four nodes of which only nodes $2$, $3$ and $4$ have a controllable current injection. The corresponding circuit is provided in Figure \ref{hvdc network}, where $\mathcal{C}_i$ is the capacitance at terminal $i$, and $\mathcal{L}_k$ is the inductance of line $k$.
The first objective is to stabilize the voltage at terminal $i$ around its desired setpoint $\overline V_i$, which is identical for each terminal. Therefore,  $B^T V = B^T (V - \overline V)$. The second objective is to share the controllable current injections equally among the terminals.
Note that (\ref{hvdcmodel}), is an example of the model studied in Subsection \ref{subpotential}, and that the set of nodes with a controllable current injection is a zero forcing set for the considered network. Therefore,
Assumption \ref{as:zero_forcing set} is satisfied and it follows from Theorem \ref{theorem3} that asymptotic stability of the desired state is guaranteed, if the controllers (\ref{controllerinput}) are applied to control the current injections.
In this case study, the controllers (\ref{controllerinput}) are applied, with $q_i =1, s_i= 0, r_i= 0$, $T_{\theta i}=100$, $T_{\phi i}=0.02$, for all $i \in \{1,2,3,4\}$. The
underlying communication network is undirected and connects nodes $2-3$ and $3-4$, where each node has a weight of $10^4$. The desired voltage is $\overline V_i = 165kV$ at all terminals throughout the simulation. Initially, all $d_i$ have a value of $100A$. At $t=0.02s$, the value of $d_2$ increased to $140A$, whereas $d_3$ is decreased to $80A$. To prevent low and high current injections during the transient we require at all terminals that $ 130 A \leq u_{i}(t) \leq 145 A$ is satisfied. To ensure this we let for all $i \in \{2,3,4\}$
\begin{align}
 u_i = g_i(\theta_i) = 130+ 7.5 \left(\tanh(\theta_i)+1\right).
\end{align}
The response to the change in demand is given in Figure \ref{case_study2}, from where we conclude that  the voltages converge towards their set point of $165kV$, while $u$ satisfies its constraints at all time.

\section{Conclusions and future directions} \label{conclusions}
We presented a distributed controller that dynamically adjusts the inputs and flows in a flow network to regulate the measured output at the nodes towards the desired value. This is achieved in presence of unknown disturbances to the network. The use of nonlinear functions, bounding the controller outputs, guarantees that the inputs and the flows stay within their capacity limits. We only require that a subset of nodes have a controllable input to obtain output regulation throughout the complete network. Additionally,
optimal coordination among the inputs, minimizing a suitable cost function, is achieved by exchanging information over a communication network. Based on Lyapunov arguments and an invariance principle, we have proven that the desired steady state is globally asymptotically attractive. We emphasized the connection to compartmental systems and we provided two case studies (a district heating system and a multi-terminal high voltage direct current network) that show the effectiveness of the proposed solution.

There are multiple interesting directions to extend the presented results. We briefly discuss a few of them. The required communication in the distributed control structure is continuous in the current setting. An interesting extension is to consider the more realistic setting where communication happens at discrete instances, leading to a hybrid system (\cite{postoyan_2015_tac}, \cite{depersis_2016_tac}). It is currently assumed that the material can be instantaneously moved from one node to another. Incorporating the possibility to include a delay in this flow is desirable (\cite{skutella2009introduction}). To cover an even larger class of physical systems, it is worthwhile to include nodes that do not have storage capabilities, which can be modelled by algebraic relations, leading to an overall algebraic-differential system, making the analysis more challenging.
Since the results are obtained without the common requirement of strict output passivity of the nodes, it is worth exploring if the proposed control structure can be applied to a wider class of systems than the considered flow networks.
\stopmodif
\section{Acknowledgement}
The authors wish to thank Pietro Tesi for his helpful comments.

\bibliographystyle{IEEEtran}
\balance
\bibliography{overallbib11}
\end{document}